\documentclass[a4paper,12pt]{article}
\pdfoutput=1
\usepackage{graphicx,rotating,hyperref,slashed,amsmath,xcolor,amssymb,amsfonts,cite,expdlist}
\makeatletter
\usepackage{soul}
\usepackage{blkarray}
\usepackage{multirow,multicol}
\usepackage{dsfont}
\usepackage{footnote}
\newcommand{\via}[1]{#1}

\makeatletter
\usepackage[natbibapa]{apacite}
\newcommand{\refchange}[2]{#2}
\AtBeginDocument{%
  
}

\hypersetup{colorlinks,bookmarksopen,bookmarksnumbered,
linkcolor=blus,pdfstartview=FitH,urlcolor=rossos,citecolor=verde}
\allowdisplaybreaks

\newcommand{\Nobel}{{\includegraphics[width=1.7ex]{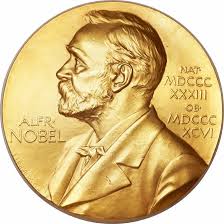}}}
\newcommand{\DiracICTP}{{\includegraphics[width=1ex]{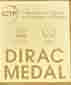}}}
\newcommand{\DiracIOP}{{\includegraphics[width=1.2ex]{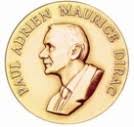}}}
\newcommand{\Milner}{{\includegraphics[width=1.2ex]{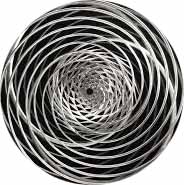}}}
\newcommand{\Planck}{{\includegraphics[width=1.2ex]{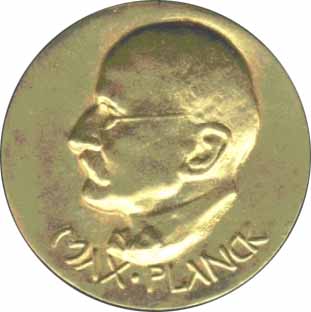}}}
\newcommand{\Sakurai}{{$\color{red}\bullet$}}
\newcommand{\Wolf}{{\includegraphics[width=1.2ex]{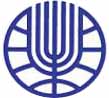}}}
\def\lsim{\mathrel{\rlap{\lower3pt\hbox{\hskip0pt$\sim$}}
   \raise1pt\hbox{$<$}}}         
\def\gsim{\mathrel{\rlap{\lower4pt\hbox{\hskip1pt$\sim$}}
   \raise1pt\hbox{$>$}}}         

\newcommand{\fig}[1]{Fig.~\ref{fig:#1}}
\newcommand{\Fig}[1]{Figure~\ref{fig:#1}}
\newcommand{\tabl}[1]{Table~\ref{tab:#1}}
\newcommand{\Tabl}[1]{Table~\ref{tab:#1}}
\newcommand{\secc}[1]{Section~\ref{sec:#1}}

\newcommand{\app}[1]{Appendix~\ref{app:#1}}

\newcommand{\eq}[1]{Eq.~\eqref{eq:#1}}

\newcommand{\reff}[1]{Ref.~\cite{#1}}
\newcommand{\reffs}[1]{Ref.s~\cite{#1}}

\newcommand{\CC}{C\hspace{-1.7ex}/\!\!/} 

\newcommand{\mio}[1]{}

\definecolor{Gray}{gray}{0.95}
\newcommand{\bbox}[1]{\fcolorbox{gray}{Gray}{~$\displaystyle #1$~}}

\definecolor{rosino}{cmyk}{0,0.05,0.05,0.02}
\definecolor{celestino}{cmyk}{0.05,0,0,0.01}

\usepackage{multicol}
\usepackage{color}
\definecolor{rosso}{cmyk}{0,1,1,0.4}
\definecolor{rossos}{cmyk}{0,1,1,0.55}
\definecolor{rossoc}{cmyk}{0,1,1,0.2}
\definecolor{blu}{cmyk}{1,1,0,0.3}
\definecolor{blus}{cmyk}{1,1,0,0.6}
\definecolor{bluc}{cmyk}{1,1,0,0.1}
\definecolor{verde}{cmyk}{0.92,0,0.59,0.25}
\definecolor{verdec}{cmyk}{0.92,0,0.59,0.15}
\definecolor{verdes}{cmyk}{0.92,0,0.59,0.4}

\def\circa#1{\,\raise.3ex\hbox{$#1$\kern-.75em\lower1ex\hbox{$\sim$}}\,}

\newcommand{\beq}{\begin{equation}}
\newcommand{\eeq}{\end{equation}}

\newcommand{\bea}{\begin{eqnarray}}
\newcommand{\eea}{\end{eqnarray}}
\newcommand{\be}{\begin{equation}}
\newcommand{\ee}{\end{equation}}
\font\tenrsfs=rsfs10 at 12pt
\font\sevenrsfs=rsfs7 at 9pt
\font\fiversfs=rsfs5
\newfam\rsfsfam
\textfont\rsfsfam=\tenrsfs
\scriptfont\rsfsfam=\sevenrsfs
\scriptscriptfont\rsfsfam=\fiversfs
\def\mathscr#1{{\fam\rsfsfam\relax#1}}

\def\circa#1{\,\raise.3ex\hbox{$#1$\kern-.75em\lower1ex\hbox{$\sim$}}\,}
\makeatletter

\def\hhref#1{\href{http://arxiv.org/abs/#1}{arXiv:#1}} 

\newcommand{\inspiref}[1]{\href{https://inspirehep.net/author/profile/#1}{#1}}

\newcommand{\doi}[1]{\href{http://dx.doi.org/#1}{[link]}}

\def\hhref#1{\href{http://arxiv.org/abs/#1}{arXiv:#1}} 
 
\def\art{\@ifnextchar[{\eart}{\oart}}
\def\eart[#1]#2#3#4#5#6{{\rm #2}, {\em #3 \bf #4} {\rm (#6) #5} ({\em #1})}

\def\article{\@ifnextchar[{\earticle}{\oarticle}}
\def\oarticle#1#2#3#4#5#6{{\rm #1}, {\em ``#6''}, {\rm #2 #3 (#5) #4}}
\def\earticle[#1]#2#3#4#5#6#7{{\rm #2}, {\em ``#7''}, {\rm #3 #4 (#6) #5}  [\hhref{#1}]}
\def\hepart[#1]#2{{\rm #2, \em#1}}
\def\heparticle[#1]#2#3{#2, {\em ``#3''} [\hhref{#1}]}
\def\inspire{{\sc InSpire} }

\newcounter{alphaequation}[equation]
\def\thealphaequation{\theequation\hbox to
0.6em{\hfil\alph{alphaequation}\hfil}}
\def\eqnsystem#1{
\def\@eqnnum{{\rm (\thealphaequation)}}
\def\@@eqncr{\let\@tempa\relax \ifcase\@eqcnt \def\@tempa{& & &} \or
  \def\@tempa{& &}\or \def\@tempa{&}\fi\@tempa
  \if@eqnsw\@eqnnum\refstepcounter{alphaequation}\fi
\global\@eqnswtrue\global\@eqcnt=0\cr}
\refstepcounter{equation} \let\@currentlabel\theequation \def\@tempb{#1}
\ifx\@tempb\empty\else\label{#1}\fi
\refstepcounter{alphaequation}
\let\@currentlabel\thealphaequation
\global\@eqnswtrue\global\@eqcnt=0 \tabskip\@centering\let\\=\@eqncr
$$\halign to \displaywidth\bgroup \@eqnsel\hskip\@centering
$\displaystyle\tabskip\z@{##}$&\global\@eqcnt\@ne
\hskip2\arraycolsep\hfil${##}$\hfil& \global\@eqcnt\tw@\hskip2\arraycolsep
$\displaystyle\tabskip\z@{##}$\hfil
\tabskip\@centering&\llap{##}\tabskip\z@\cr}
\def\endeqnsystem{\@@eqncr\egroup$$\global\@ignoretrue} \makeatother

\definecolor{fiorentina}{rgb}{.5,0,.5}

\definecolor{rossoCP3}{cmyk}{0,.88,.77,.40}

\begin{document}
\centerline{\hfill CERN-TH-2018-066}

\vspace{2cm}

\begin{center}
\boldmath

{\textbf{\LARGE\color{rossoCP3} Biblioranking fundamental physics}}\\[4mm]
{\textbf{\LARGE\color{rossoCP3} (updated to 2021/1/1)}}
\unboldmath

\bigskip\bigskip

\vspace{0.1truecm}

{\large\bf Alessandro Strumia$^{a}$, Riccardo Torre$^{b,c}$}
 \\[8mm]
{\it $^a$ Dipartimento di Fisica dell'Universit{\`a} di Pisa, Italy}\\[1mm]
{\it $^b$ CERN, Theory Division, Geneva, Switzerland}\\[1mm]
{\it $^c$ INFN, sezione di Genova, Italy}\\[1mm]

\vspace{1cm}

\thispagestyle{empty}
{\large\bf\color{blus} Abstract}
\begin{quote}
\large
Counting of number of papers, of citations and the $h$-index
are the simplest bibliometric indices of the impact of research.
We discuss some improvements.
First, we replace citations with {\em individual citations},  fractionally shared among co-authors,
to take into account that different papers and different fields have largely different average number of co-authors and of references. 
Next, we improve on citation counting applying the PageRank algorithm to citations among papers. 
Being time-ordered, this reduces to a weighted counting of citation descendants that we call {\it PaperRank}. 
We compute a related {\it AuthorRank} applying the PageRank algorithm to citations among authors.
These metrics quantify the impact of an author or paper taking into account the impact of those authors that cite it. 
Finally, we show how self- and circular- citations can be eliminated by defining
a closed market of {\it Citation-coins}.
We apply these metrics to the \inspire database that covers fundamental physics, presenting results for papers, authors, journals, institutes, towns, countries\via{, and continents,}
for all-time and in recent time periods.
\end{quote}
\thispagestyle{empty}
\end{center}

\setcounter{page}{1}
\setcounter{footnote}{0}

\newpage

\via{\tableofcontents}

\section{Introduction}
Bibliometrics can be a useful tool for evaluating research: it provides simple, quick, first objective measures of the impact of papers and authors
and is increasingly being considered a useful (although incomplete) evaluation criterion in postdoc/faculty recruitments, fundings, and grant awards \citep{2017JASIS..68..695K,2017arXiv170602153H,2017arXiv170709955K}. 
In the fundamental physics community that we consider in this paper, the most common measures of the impact such as counting of number of papers, citations and Hirsch's $h$-index \citep{Hirsch:2005zc}, are inflating \citep{Sinatra_2015}, making it harder to identify the real impact of research, especially for the most recent literature. 
The more papers one writes and the more citations these papers get, the bigger bibliometric estimators become:
it does not matter if these citations close in certain loops and/or remain confined in sub-fields, or whether the paper has been written by a single author or in collaboration with thousands of people.

We introduce new metrics and compare them with the existing ones, showing how they address the issues mentioned above. 
\refchange{
We apply them to the \inspire bibliographic database \citep{inspirewebpage,Holtkamp:2010cca,Ivanov:2010zzf,Klem:2011zz,MartinMontull:2011tya},\footnote{For other notable digital libraries and databases of research in various fields, see \reffs{adsnasa,cerncds,arxivorg,googlescholar,microsoftacademic,dblpcomputerscience,acbdiglib,pubmed,mathscinet,citeseer,semanticscholar,repec,ieeexplore,zbmath}.}}{We apply them to the \inspire\footnote{\href{https://inspirehep.net}{https://inspirehep.net}.} bibliographic database \citep{Holtkamp:2010cca,Ivanov:2010zzf,Klem:2011zz,MartinMontull:2011tya},\footnote{Other notable digital libraries and databases of research literature in various fields are, for instance, ADS - The SAO/NASA Astrophysics Data System (\href{http://www.adsabs.harvard.edu}{http://www.adsabs.harvard.edu}), CDS - CERN Document Server (\href{https://cds.cern.ch}{https://cds.cern.ch}), arXiv.org (\href{https://arxiv.org}{https://arxiv.org}), Google Scholar (\href{https://scholar.google.it}{https://scholar.google.it}), Microsoft Academic (\href{https://academic.microsoft.com}{https://academic.microsoft.com}), DBLP - Computer Science Bibliography (\href{https://dblp.uni-trier.de}{https://dblp.uni-trier.de}), ACM Digital Library (\href{https://dl.acm.org}{https://dl.acm.org}), PubMed (\href{https://www.ncbi.nlm.nih.gov/pubmed}{https://www.ncbi.nlm.nih.gov/pubmed}), MathSciNet - Mathematical Reviews (\href{https://mathscinet.ams.org}{https://mathscinet.ams.org}), CiteSeerX (\href{https://citeseerx.ist.psu.edu}{https://citeseerx.ist.psu.edu}), Semantic Scholar (\href{https://www.semanticscholar.org}{https://www.semanticscholar.org}), RePEc - Research Papers in Economics (\href{http://repec.org}{http://repec.org}), IEEE Xplore Digital Library (\href{https://ieeexplore.ieee.org}{https://ieeexplore.ieee.org}), and zbMATH the first resource for mathematics (\href{https://zbmath.org}{https://zbmath.org}).}} 
that covers fundamental physics literature after $\approx1970$.\footnote{The most relevant literature before $\approx 1970$ has been added and is still being added on a request base to {\sc InSpire}.} The metrics, both the usual ones and the new ones that we introduce, can measure the impact of papers, $p,p',\ldots$, of authors, $A,A',\ldots$, and of groups. They are defined as follows:

\begin{enumerate}
\item {\bf Number of papers}\\
The most naive metric consists in counting the number of papers $N^{\rm pap}_A= \sum_{p\in A}1$ 
written by a given author $A$. 
This metric rewards the most prolific authors.

\item {\bf Number of citations}\\
The most used metric consists in counting the number of citations
$N^{\rm cit}_p$ received by a paper $p$.
An author $A$ is then evaluated summing the number of citations $N^{\rm cit}_A$
received by its  papers.
In formul\ae:
\beq\label{eq:usual}
N^{\rm cit}_p = \sum_{p'\to p}1\,,\qquad
N^{\rm cit}_A = \sum_{p\in A} N^{\rm cit}_p\,,\eeq
where the first sum runs over all papers $p'$ that cite $p$, and the second sum over all papers $p$ of author $A$.

\item {\bf $h$-index}\\
The $h$-index is defined as the maximum $h$ such that $h$ papers have at least $h$ citations each. 
In formul\ae, assuming that all papers of author $A$ are sorted in decreasing order of number of citations $N^{\rm cit}_{p}\geq N^{\rm cit}_{p+1}$, it is given by
\beq\label{eq:average}
h_A=\max \left\{p \,|\, p \leq N^{\rm cit}_{p}\right\}\,.
\eeq
This is proportional to $\sqrt{N^{\rm cit}_A}$, times a factor that
penalises authors that write a small number of highly cited papers \citep{Hirsch:2005zc}.

\smallskip

\listpart{As we will see, the average number of authors per paper and of references per paper increased, in the last 20 years, by one to a few per-cent a year, and is significantly different in different communities. Following basic common-sense, we propose an improved metric that renormalises away such factors,
and that cannot be artificially inflated adding more references and/or more co-authors.}

\item {\bf Number of individual citations}\\
A citation from paper $p'$ to paper $p$ is weighted as the inverse of the number of references $N_{p'}^{\rm ref}$ of paper $p'$. Furthermore, the citations received by a paper $p$ are equally shared among its $N_p^{\rm aut}$ authors.\footnote{We assume that authors contributed equally
because in fundamental physics authors are usually listed alphabetically, with no information about who contributed more.
The factor $1/N_p^{\rm aut}$, that is the one dictated by conservation laws, will be further motivated in \secc{FC}.}
In formul\ae:
\beq\label{eq:CA}\bbox{
N^{\rm icit}_p = \sum_{p'\to p}\frac{1}{N_{p'}^{\rm ref}}\,,\qquad
N^{\rm icit}_A = \sum_{p\in A} \frac{N^{\rm icit}_p}{N_p^{\rm aut}}}\,.\eeq
\begin{savenotes}
\listpart{
$\quad\,$  The definition of individual quantities is not new to the scientometrics literature and has been extensively studied as an application of the so-called {\it fractional counting}. Fractional counting has been considered both in the context of metrics and rankings \citep{DBLP:journals/jasis/Hooydonk97,DBLP:journals/jasis/Egghe08a,DBLP:journals/corr/abs-1007-4749,DBLP:journals/corr/abs-1006-2896,DBLP:journals/corr/abs-1106-0114,DBLP:journals/jasis/LeydesdorffS11,DBLP:journals/jasis/LeydesdorffB11,DBLP:journals/joi/AksnesSG12,DBLP:journals/aslib/Rousseau14,DBLP:journals/joi/BouyssouM16}, and in the context of constructing research networks \citep{DBLP:journals/joi/Perianes-Rodriguez16,DBLP:journals/joi/LeydesdorffP17}.

$\quad\,$ Furthermore, as well known in the literature,
the division by $N_{p'}^{\rm ref}$ factors out the different
publication intensity in the various fields, without the need of a field classification system.
Indeed, a paper has $N_p^{\rm icit}=1$ if it receives the mean number of citations in its field.

\smallskip

$\quad\,$ All the above metrics are defined ``locally'', i.e.~they can be computed for a paper/author without knowing anything about all other papers but the ones that cite it. Therefore, they all potentially suffer from the problem that even small sub-communities can  inflate their indicators. To overcome this problem one needs to define global measures, i.e.~measures that know about the whole community. The simplest such global measure of impact is given by the {\it PageRank} algorithm, introduced in 1996 by Larry Page and Sergey Brin,\footnote{A similar idea, applied to bibliometrics, have been proposed long before the advent of the PageRank \refchange{in \reff{pinski1976citation}}{by \cite{pinski1976citation}}.} the founders of Google \citep{brinpage1998bp,brinpage1999bp}.\footnote{Even if the word ``Page'' in PageRank may seem to refer to webpages, the name of the algorithm originates from the name of one of its inventors, Larry Page.} For a pedagogical introduction to the PageRank see \refchange{\reffs{Rajaraman_2009,tanaseradu}}{\protect\cite{Rajaraman_2009} and \protect\cite{tanaseradu}}. Applications of the PageRank algorithm to citations network have already been considered, for instance, \refchange{in \reffs{Chen:2006fs,DBLP:journals/ipm/MaGZ08,DBLP:journals/jasis/DingYFC09,DBLP:conf/icdm/ZhouOZG07}}{by \cite{Chen:2006fs,DBLP:journals/ipm/MaGZ08,DBLP:journals/jasis/DingYFC09}, and \cite{DBLP:conf/icdm/ZhouOZG07}}.  More advanced ranking algorithms, based on integrated bibliographic information have also been proposed \refchange{in \reffs{zbMATH05602132,Bini2010ACA,zbMATH05644252}}{by \cite{zbMATH05602132,Bini2010ACA}, and \cite{zbMATH05644252}}.}
\end{savenotes}

\item {\bf PaperRank}\\
The PaperRank $R_p$ of paper $p$ and the PaperRank $R_A$ of author $A$ are defined as
\beq\label{eq:ranks}\bbox{
R_p = \sum_{p'\to p}\frac{R_{p'}}{N_{p'}^{\rm ref}}\,,\qquad
R_A = \sum_{p\in A} \frac{R_p}{N_p^{\rm aut}}}\,.\eeq
Namely, citations from papers $p'$ are weighted proportionally to their ranks $R_{p'}$,
that get thereby determined trough a system of linear equations.
The PaperRank provides a metric which cannot be easily artificially inflated,
because it is the bibliometric estimator of a physical quantity: 
how many times each paper is read.
\begin{savenotes}
\listpart{As we will see, the PaperRank singles out notable old papers which often do not have many citations.
However, given that citations are time-ordered (newer papers cite older ones), 
the rank reduces to a weighted sum over citation descendants (a combination of ``citations of citations'')
which needs about 10-20 years to become a better indicator than the number of individual citations.
In order to use information from the past, we define an alternative AuthorRank based on citations among authors.}
\end{savenotes}

\item {\bf AuthorRank}\\
We define the citation matrix which counts all individual citations from
author $A'$ to $A$
\beq \label{eq:Nicitmatrix}
N^{\rm icit}_{A'\to A} = \sum_{p_{A'} \to p_{A} } \frac{1}{N_{p_A}^{\rm aut}  N_{p_{A'}}^{\rm aut}}\frac{1}{N_{p_{A'}}^{\rm ref}}\,,
\eeq
where the sum runs over all papers $p_{A'}$ of author $A'$ that cite papers $p_{A}$ of $A$.
We then define the AuthorRank $\mathscr{R}_A$ as
\beq \label{eq:Cmatrix}
\bbox{\mathscr{R}_A = \sum_{A'} \mathscr{R}_{A'} C_{A'\to A}\,,\qquad
C_{A'\to A}= \frac{N^{\rm icit}_{A'\to A}}{\sum_{A''} N^{\rm icit}_{A'\to A''}}}\,,
\eeq
namely, as the principal eigenvector of the right stochastic matrix $C_{A'\to A}$,
which (thanks to the normalization of each row provided by the sum in the denominator) tells the percentage of individual citations to $A$ among all individual citations of $A'$.
The AuthorRank gives more weight to citations coming from highly cited authors. We also use the AuthorRank of authors to define an improved ranking of papers, that we call AuthorRank of papers, as
\beq \label{eq:ARp}
\bbox{ \mathscr{R}_p =   \sum_{p'\to p} \sum_{A\in p'} \frac{\mathscr{R}_{A}}{N_{p'}^ {\rm aut}N_{p'}^{\rm ref}}}\,.\eeq

\end{enumerate} 
{Ideas similar to our AuthorRank have already been considered in the literature \citep{DBLP:journals/corr/abs-0907-1050,DBLP:journals/jasis/WestJDGB13}.  
\refchange{The authors of \reff{DBLP:journals/corr/abs-0907-1050}}{\cite{DBLP:journals/corr/abs-0907-1050}}
applied an algorithm similar to our AuthorRank to the Physical Review publication archive up to 2006.
The implementation, called Science Author Rank Algorithm (SARA) is different than ours in few respects: first, it is based on slices of the full database, given by multiple graphs with equal number of citations, while our AuthorRank includes information from a single graph constructed with weighted citations among all authors after a given time (or for all times). Second, the SARA authors consider about half the number of papers we consider and around one third of the citations we consider, all coming from the single publisher Physical Review. While the analysis of \refchange{\reff{DBLP:journals/corr/abs-0907-1050}}{\cite{DBLP:journals/corr/abs-0907-1050}} is pioneering in this direction, we believe that focusing on a single publisher introduces bias, since it does not cover the whole scientific production of scientists: only a fraction of each author's papers are published on a Physical Review journal. We avoid this bias by considering the \inspire public database including the scientific research in Fundamental Physics consisting of preprints and articles published on about $2\cdot 10^{3}$ journals. This should remove some of the bias induced by considering a single publisher. }

{Third and most important difference between SARA and the AuthorRank is the contribution of citations among papers to the weight of the links in the author-level graph: in the case of SARA each paper contributes to the author-level graph with a weight $1/(m\cdot n)$, with $m$ and $n$ the number of authors in the citing and cited paper. In our case instead this weight is $1/N^{\text{icit}}_{A'\to A} $, with this quantity defined in \eq{Nicitmatrix}. As can be seen by this equation, this also contains a $1/N_{p_{A'}}^{\rm ref}$. In simpler words, the SARA author-level graph is determined using citations among papers, in our notation $N^{\rm cit}_p$, while our author-level graph is determined using individual citations among papers, in our notation $N^{\rm icit}_p$, as defined in \eq{CA}. We find the latter more indicative, as we find individual citations of papers, that are the building blocks of the paper-level graph, more indicative than traditional citations. Being cited among many others is typically less relevant than being cited with a few other references.

\refchange{The authors of }{}\cite{DBLP:journals/jasis/WestJDGB13} applied a similar algorithm to the Social Science Research Network Community, calling it Author-Level Eigenfactor Metric. In this case some bias is introduced by removing all self-citations from the graph, corresponding to the diagonal of the correlation matrix $N_{A'\to A}^{\text{icit}}$. As we discuss in \secc{cartels}, removing self-citations is an arbitrary procedure, also admitting different implementations. The problem of making the effect of self citations less relevant can be addressed in two different ways: by tuning the teleport probability parameter in the PageRank algorithm, or by considering a more general quantity, like the Citation-coin introduced below.

\indent In this paper we stress that, while the AuthorRank is an interesting metric by itself, it mainly identifies already well known authors, like Feynman and others \citep{DBLP:journals/corr/abs-0907-1050}, and Jensen and others \citep{DBLP:journals/jasis/WestJDGB13}. We try to improve in extracting information from the AuthorRank and use it to identify recent papers cited by highly ranked (with AuthorRank) physicists. This allows us to define an AuthorRank for papers, that is a good candidate as an early alert for potentially important papers.

As we stated before, removing self-citations carries some level of arbitrariness. One could aim at resolving a better defined problem, consisting of removing all citation `cartels', defined as loops of citations among 3 authors, among 4 authors, etc.
The mathematical problem of removing all circular citations has a simple solution, inspired by economy: money.\footnote{In mathematical language this consists in removing all closed loops from the authors' graph, making it acyclic.}
You don't get richer by giving money to yourself or  by circulating money with friends. This leads us to the definition of an additional metric:}

\begin{enumerate}
\item[7.] {\bf Citation-coin}\\
Author $A$ `owes' the number $\CC_A$ of individual citations received minus the number of individual citations given:\footnote{A slight improvement will be added later, to avoid border effects due to intrinsic finite nature of the database.}
\beq  \bbox{
\CC_A =\sum_{A'} (N^{\text{icit}}_{A'\to A}-N^{\text{icit}}_{A\to A'})= N^{\rm icit}_A - \sum_{p\in A} \frac{1}{N_p^{\rm aut}}}\,.
\eeq
This metric penalises authors who write many small papers which receive few citations from others.
\end{enumerate} 
An important property of all above metrics is that they can be computed in practice.
Our metrics are intensive: so they
can be used to rank groups, such as journals, institutes, countries, etc.\ by simply summing over their members.
Furthermore they can be restricted to any specific time period, e.g.\ after year 2000.

\medskip

The paper is structured as follows.
In \secc{PaperRank} we introduce the
PaperRank $R$, discuss its features and properties, and rank all papers in {\sc InSpire}.
In \secc{AuthorRank} we  introduce the number of individual citations $N_{\rm icit}$, the AuthorRank $\mathscr{R}$
and the Citation-coin $\CC$,
discussing their features and properties, and rank all authors in {\sc InSpire}.
In \secc{GeoRank} we apply these measures to rank groups:
institutions, towns, countries, \via{continents,} and journals.
Conclusions are presented in \secc{concl}. 
In \app{InSpire} we describe the \inspire and arXiv databases and their main features and trends,
together with technical details.

Several of our results with complete tables are available at the PhysRank \refchange{webpage~\cite{webpage}}{webpage\footnote{\href{http://rtorre.web.cern.ch/rtorre/PhysRank}{http://rtorre.web.cern.ch/rtorre/PhysRank}.}}.

\section{Ranking papers}\label{sec:PaperRank}

\subsection{PaperRank}\label{sec:rank}
Given a citation network of $N_{\rm pap}$ papers, the PaperRank $R_p$ of each paper $p$ is defined by 
\beq
R_p = \wp \sum_{p'\to p} \frac{R_{p'}}{N_{p'}^{\rm ref}}+ \alpha (1-\wp)\,,
\label{eq:google}
\eeq
where $N_{p'}^{\rm ref}$ is the total number of references of each paper $p'$ that cites $p$.
Equation\eq{google} is linear, so its solution is unique, and can be computed efficiently iteratively \citep{Rajaraman_2009}.

We elaborate on its meaning.
Eq.\eq{google} contains two arbitrary constants $\alpha$ and $\wp$.
The constant $\alpha$ just fixes the overall normalization $\sum_p R_p = R_{\rm tot}$.
When applied to internet, $R_p$ describes the probability that site $p$ is visited and
it is convenient to normalize it to one.
We choose $R_{\rm tot}$ equal to the total number of citations $ R_{\rm tot}=\sum_{p}N_{p}^{\text{cit}}$, in order to allow for an easier
comparison between the number of citations received by a paper and its PaperRank $R_p$.
With this normalisation $R_p$ grows with time, as newer papers appear.

Viewing the rank as the probability that a paper is read, 
the parameter $ \wp $ splits it into two contributions:
the first term is the probability that a reader reaches a paper by following a reference to it;
the second term, equal for all papers, simulates readers that randomly browse the literature.
\begin{itemize}
\item In the limit $\wp=0$ the first contribution vanishes, and all papers have a common rank.
At first order in small $\wp\ll 1$, $R_p$ starts discriminating the papers $p$:
\beq  \label{eq:Cp}
R_p (\wp) \stackrel{\wp\ll 1}{\simeq}  \frac{R_{\rm tot}}{N_{\rm pap}}
\frac{1+ \wp N^{\rm icit}_p}{1+\wp}\,,
\eeq
where $N^{\rm icit}_p$ is the number of individual citations received by $p$ defined in \eq{CA}, which 
obeys $\sum_p N^{\rm icit}_p = N_{\rm pap}$.

\item 
In the limit $\wp=1$ the second contribution in \eq{google} vanishes, and $R_p$ only depends on the structure of the network,
provided that no closed sub-networks and dead-ends exist~\citep{Rajaraman_2009}.
Recursive computations of $R_p$ become slower as $\wp\to 1$.
\end{itemize}
Data about downloads of scientific articles would allow to extract the value of $\wp$ that better fits the observed reading rate;
however such data are not available in fundamental physics.\footnote{arXiv.org does not make public the number of downloads, to avoid the conversion of this information into a relevant metric, and its consequent fate determined by Goodhart's law.} We use a large $\wp=0.99$, such that the first contribution in \eq{google} dominates for all relevant authors.

\begin{table}  \renewcommand{\arraystretch}{0.9}\setlength\tabcolsep{3pt}
\begin{center}\small
\begin{tabular}{cllcc|ccc}
 & Title & 1st author & $N_{\rm aut}$ & date  & \color{blue}$N_{\rm cit}$ & $R_p$ & $\mathscr{R}_p$  \\  \hline 
1 & \href{http://inspirehep.net/record/451647}{\em The Large N limit of superconformal fiel} & J.M.Maldacena & 1 & 1998 & \color{blue}16317 & 6725 & 22034\\ 
2 & \href{http://inspirehep.net/record/593382}{\em GEANT4--a simulation toolkit} & James.R.Allison & 127 & 2003 & \color{blue}13488 & 6790 & 2170\\ 
3 & \href{http://inspirehep.net/record/484837}{\em Measurements of $\Omega$ and $\Lambda$ f} & S.Perlmutter & 32 & 1999 & \color{blue}12983 & 4124 & 7543\\ 
4 & \href{http://inspirehep.net/record/470671}{\em Observational evidence from supernovae f} & A.G.Riess & 20 & 1998 & \color{blue}12965 & 4267 & 7376\\ 
5 & \href{http://inspirehep.net/record/51188}{\em A Model of Leptons} & Steven.Weinberg & 1 & 1968 & \color{blue}12955 & 44908 & 41591\\ 
6 & \href{http://inspirehep.net/record/712925}{\em PYTHIA 6.4 Physics and Manual} & T.Sjostrand & 3 & 2006 & \color{blue}11823 & 3439 & 3798\\ 
7 & \href{http://inspirehep.net/record/1124337}{\em Observation of a new particle in the sea} & ATLAS & 2932 & 2012 & \color{blue}11666 & 1652 & 4074\\ 
8 & \href{http://inspirehep.net/record/1124338}{\em Observation of a New Boson at a Mass of } & CMS & 2897 & 2012 & \color{blue}11404 & 1638 & 4014\\ 
9 & \href{http://inspirehep.net/record/81350}{\em CP Violation in the Renormalizable Theor} & M.Kobayashi & 2 & 1973 & \color{blue}10695 & 10903 & 20359\\ 
10 & \href{http://inspirehep.net/record/467400}{\em Anti-de Sitter space and holography} & E.Witten & 1 & 1998 & \color{blue}10513 & 4471 & 13998
\end{tabular}
\caption{\em\label{tab:topcited0} Top-cited (highest number of citations) papers in the \inspire database.}
\vspace{0.5cm} \begin{tabular}{cllcc|ccc}
 & Title & 1st author & $N_{\rm aut}$ & date  & $N_{\rm cit}$ & \color{blue} $R_p$ & $\mathscr{R}_p$  \\  \hline 
1 & \href{http://inspirehep.net/record/51188}{\em A Model of Leptons} & Steven.Weinberg & 1 & 1968 & 12955 & \color{blue}44908 & 41591\\ 
2 & \href{http://inspirehep.net/record/12285}{\em Conservation of Isotopic Spin and Isotop} & C.N.Yang & 2 & 1954 & 2852 & \color{blue}41770 & 14876\\ 
3 & \href{http://inspirehep.net/record/1115221}{\em Theory of Fermi interaction} & R.P.Feynman & 2 & 1958 & 1748 & \color{blue}39446 & 10778\\ 
4 & \href{http://inspirehep.net/record/1299937}{\em Remarks on the Dirac theory of the posit} & W.Heisenberg & 1 & 1934 & 122 & \color{blue}38682 & 1223\\ 
5 & \href{http://inspirehep.net/record/9146}{\em On the Stopping of fast particles and on} & H.A.Bethe & 2 & 1934 & 724 & \color{blue}31585 & 1009\\ 
6 & \href{http://inspirehep.net/record/9087}{\em The S matrix in quantum electrodynamics} & F.J.Dyson & 1 & 1949 & 771 & \color{blue}31296 & 5044\\ 
7 & \href{http://inspirehep.net/record/2477}{\em Symmetries of baryons and mesons} & M.Gell.Mann & 1 & 1962 & 1666 & \color{blue}29947 & 8576\\ 
8 & \href{http://inspirehep.net/record/12289}{\em Field Theories with Superconductor Solut} & J.Goldstone & 1 & 1961 & 2078 & \color{blue}29244 & 5773\\ 
9 & \href{http://inspirehep.net/record/12286}{\em A Theory of the Fundamental Interactions} & J.S.Schwinger & 1 & 1957 & 679 & \color{blue}28331 & 11962\\ 
10 & \href{http://inspirehep.net/record/1115218}{\em Space - time approach to quantum electro} & R.P.Feynman & 1 & 1949 & 830 & \color{blue}26963 & 3550
\end{tabular}
\caption{\em\label{tab:topranked0} Top-ranked (highest PaperRank) papers in the \inspire database.}
\vspace{0.5cm}\begin{tabular}{cllcc|ccc}
 & Title  & 1st author & $N_{\rm aut}$ & date  & $N_{\rm cit}$& $R_p$ & \color{blue}$\mathscr{R}_p$ \\  \hline 
1 & \href{http://inspirehep.net/record/51188}{\em A Model of Leptons} & Steven.Weinberg & 1 & 1968 & 12955 & 44908 & \color{blue}41591\\ 
2 & \href{http://inspirehep.net/record/101338}{\em Particle Creation by Black Holes} & S.W.Hawking & 1 & 1974 & 8665 & 9142 & \color{blue}28472\\ 
3 & \href{http://inspirehep.net/record/80491}{\em A Planar Diagram Theory for Strong Inter} & G.tHooft & 1 & 1974 & 4989 & 8238 & \color{blue}26994\\ 
4 & \href{http://inspirehep.net/record/92111}{\em Unity of All Elementary Particle Forces} & H.M.Georgi & 2 & 1974 & 5090 & 10397 & \color{blue}25424\\ 
5 & \href{http://inspirehep.net/record/89145}{\em Confinement of Quarks} & K.G.Wilson & 1 & 1974 & 5376 & 23570 & \color{blue}23192\\ 
6 & \href{http://inspirehep.net/record/60999}{\em Weak Interactions with Lepton-Hadron Sym} & S.L.Glashow & 3 & 1970 & 6214 & 19199 & \color{blue}22997\\ 
7 & \href{http://inspirehep.net/record/101400}{\em Pseudoparticle Solutions of the Yang-Mil} & A.A.Belavin & 4 & 1975 & 2887 & 13352 & \color{blue}22315\\ 
8 & \href{http://inspirehep.net/record/451647}{\em The Large N limit of superconformal fiel} & J.M.Maldacena & 1 & 1998 & 16317 & 6725 & \color{blue}22034\\ 
9 & \href{http://inspirehep.net/record/81350}{\em CP Violation in the Renormalizable Theor} & M.Kobayashi & 2 & 1973 & 10695 & 10903 & \color{blue}20359\\ 
10 & \href{http://inspirehep.net/record/3438}{\em Symmetry Breaking Through Bell-Jackiw An} & G.tHooft & 1 & 1976 & 3769 & 8220 & \color{blue}20081
\end{tabular}
\caption{\em\label{tab:topreferred0} Top-referred (highest AuthorRank) papers in the \inspire database.}
\end{center}\normalsize
\end{table}

\subsection{PaperRank of papers: results}
We compute the PaperRank by constructing a graph (and its transition matrix) having all papers as nodes and citations as links. We consider the full \inspire database, as detailed in \app{ArXiv1}. Generally, a few hundred iterations of \eq{google} are necessary for a percent level convergence. 
The computation takes a few minutes on a laptop computer.

\Tabl{topcited0} (\tabl{topranked0}) shows the top-cited (top-ranked) papers
in the \inspire database.
Top-ranked papers correspond to the papers with top PaperRank and tend to be old famous ones, even with a relatively small number of citations.
Top-cited papers, ranked with the usual counting of the number of citations, tend to be modern, in the view of the inflation in the rate of citations.
The same effect was observed \refchange{in \reff{Chen:2006fs}}{by \cite{Chen:2006fs}}, who applied the PageRank algorithm to the sub-set of
papers published on Physical Review.

The difference between the two rankings is partly due to the fact that PaperRank penalises recent papers.
Papers tend to accumulate citations for about 10-20 years, while the rank continues growing with time,
and is highly suppressed for younger papers.

\medskip

This also means that the PaperRank defined in \eq{google} needs 10-20 years before providing a better metrics than the number of citations.
This is proven in the next section, where we show that, for a time-ordered network (such as the network of citations), the PaperRank reduces to the number of citations-of-citations.

\subsection{PaperRank as the number of citations-of-citations}\label{sec:Rtheo}
Internet allows for reciprocal links among pages, and the PageRank captures in a simple way the self-interacting system.
Citations among scientific papers are instead time-ordered, forming an acyclic network.
In the limit where citations of older papers to newer papers are ignored,\footnote{We enforced time-ordering within the citations,
deleting from the \inspire database a small number of `a-causal' citations, where older papers cite newer papers (see \app{ArXiv1} for details).
Since older papers tend to accumulate large ranks, a-causal citations can artificially inflate the rank of a few recent papers.}
 no loops are possible within the network, and 
the implicit definition of the rank $R_p$ of \eq{google} can be converted into the following explicit expression\footnote{This can be proven by substituting
\eq{explicit} into \eq{google}. 
A physicist can view in \eq{explicit} a path-integral within the network.
}
\beq \label{eq:explicit}{
R_p \propto \sum_{g=0}^\infty  \wp^g \sum_{p_g\to \cdots\to p}\frac{1}{N_{p_g}^{\rm ref}}\cdots \frac{1}{N_{p_1}^{\rm ref}}}\,.\eeq
Basically, $R_p$ counts the number of citations-of-citations up to generation $g$.
In the above expression, the term with
\begin{itemize}
\item[-] $g=0$ contributes as unity, and
accounts for the constant term in \eq{google}, which is negligible for papers that receive citations from others;
\item[-]  $g=1$ contributes with the number of individual citations $N^{\rm icit}_p$ as in \eq{Cp}: the sum runs over
`first generation' papers $p_1$ that cite the paper $p$;
\item[-]  $g=2$ corresponds to `second generation' papers $p_2$ that cite the papers $p_1$ that cite $p$;
\item[-]  $g=3$ corresponds to `third generation' papers $p_3$ that cite the papers $p_2$ that cite the papers $p_1$ that cite $p$;
\item[-] for generic $g$ the sum runs over all papers $p_g$ that cite paper $p$ in $g$ steps.
\end{itemize}
In other terms, for any given paper, we refer to papers that cite it as ``first generation'',
and  define as ``second generation'' those papers that cite at least a first generation paper, and so on.
A paper $q$ can appear multiple times in different generations $g$, corresponding to all possible citations paths from $q$ to $p$.
Eq.\eq{explicit} shows that $\wp<1$ gives a cut-off on the number of generations that one wants to consider,
and that $R_p(\wp)$ grows with $\wp$, and with time.
\begin{itemize}
\item At one extremum, $\wp\to 0$, the rank $R_p$ reduces to the ``number of children'' $N^{\rm icit}_p$,
without checking if they are successful.
Papers on hot topics can fast accumulate many citations,
even if later the hot topic becomes a dead topic.
Too recent papers are penalised.

\begin{figure}
$$\includegraphics[width=0.95\textwidth]{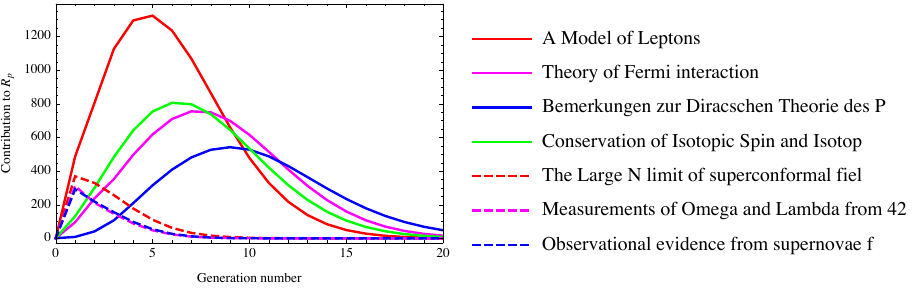}$$
\caption{\label{fig:RankGen}\em Contributions of each generation to
the citation chain of some notable papers.}
\end{figure}

\item At the other extremum, $\wp\to 1$, the rank $R_p$ becomes the Adamo number: it counts descendants.
Seminal papers that open new successful fields are rated highly,
and their rank continues to grow.
Too recent papers are highly penalised.

\end{itemize}

\begin{table}[t]
\begin{center}\small \renewcommand{\arraystretch}{0.9}\setlength\tabcolsep{2pt}
\begin{tabular}{cllc|ccc}
 Year & Title & 1st author & $N_{\rm aut}$  & $N_{\rm cit}$ & $R_p$ & $\mathscr{R}_p$  \\  \hline 
1880 & \href{http://inspirehep.net/record/1127449}{\em On the Relative Motion of the Earth and the} & A.A.Michelson & 2 & \color{blue}211 & \color{blue}219 & \color{blue}127\\ 
\hline1890 & \href{http://inspirehep.net/record/45181}{\em Cathode rays} & J.J.Thomson & 1 & \color{blue}239 & \color{blue}25 & \color{blue}33\\ 
\hline1900 & \href{http://inspirehep.net/record/4181}{\em On the electrodynamics of moving bodies} & Albert.Einstein & 1 & \color{blue}499 & \color{blue}754 & \color{blue}1272\\ 
\hline1910 & \href{http://inspirehep.net/record/4180}{\em The Foundation of the General Theory of Rel} & Albert.Einstein & 1 & \color{blue}1104 & 1415 & 973\\ 
1910 & \href{http://inspirehep.net/record/42612}{\em Approximative Integration of the Field Equa} & Albert.Einstein & 1 & 331 & 1552 & \color{blue}6501\\ 
1910 & \href{http://inspirehep.net/record/1088518}{\em Einstein's theory of gravitation and its as} & de Sitter & 1 & 219 & \color{blue}3474 & 182\\ 
\hline1920 & \href{http://inspirehep.net/record/1486}{\em About the Pauli exclusion principle} & E.P.Wigner & 2 & 383 & \color{blue}23416 & 2222\\ 
1920 & \href{http://inspirehep.net/record/14622}{\em Quantum Theory and Five-Dimensional Theory } & O.Klein & 1 & \color{blue}2481 & 3233 & \color{blue}5741\\ 
\hline1930 & \href{http://inspirehep.net/record/4677}{\em Quantised singularities in the electromagne} & P.A.M.Dirac & 1 & \color{blue}2269 & 20421 & \color{blue}12586\\ 
1930 & \href{http://inspirehep.net/record/1299937}{\em Remarks on the Dirac theory of the positron} & W.Heisenberg & 1 & 122 & \color{blue}38682 & 1223\\ 
\hline1940 & \href{http://inspirehep.net/record/1075}{\em The Theory of magnetic poles} & P.A.M.Dirac & 1 & 1071 & 6126 & \color{blue}6821\\ 
1940 & \href{http://inspirehep.net/record/9087}{\em The S matrix in quantum electrodynamics} & F.J.Dyson & 1 & 771 & \color{blue}31296 & 5044\\ 
1940 & \href{http://inspirehep.net/record/26316}{\em Forms of Relativistic Dynamics} & P.A.M.Dirac & 1 & \color{blue}1828 & 1643 & 4095\\ 
\hline1950 & \href{http://inspirehep.net/record/113}{\em On gauge invariance and vacuum polarization} & J.S.Schwinger & 1 & \color{blue}5160 & 10263 & 11589\\ 
1950 & \href{http://inspirehep.net/record/12285}{\em Conservation of Isotopic Spin and Isotopic } & C.N.Yang & 2 & 2852 & \color{blue}41770 & \color{blue}14876\\ 
\hline1960 & \href{http://inspirehep.net/record/51188}{\em A Model of Leptons} & Steven.Weinberg & 1 & \color{blue}12955 & \color{blue}44908 & \color{blue}41591\\ 
\hline1970 & \href{http://inspirehep.net/record/81350}{\em CP Violation in the Renormalizable Theory o} & M.Kobayashi & 2 & \color{blue}10695 & 10903 & 20359\\ 
1970 & \href{http://inspirehep.net/record/89145}{\em Confinement of Quarks} & K.G.Wilson & 1 & 5376 & \color{blue}23570 & 23192\\ 
1970 & \href{http://inspirehep.net/record/101338}{\em Particle Creation by Black Holes} & S.W.Hawking & 1 & 8665 & 9142 & \color{blue}28472\\ 
\hline1980 & \href{http://inspirehep.net/record/154280}{\em The Inflationary Universe: A Possible Solut} & A.H.Guth & 1 & \color{blue}8171 & \color{blue}9250 & \color{blue}15347\\ 
\hline1990 & \href{http://inspirehep.net/record/451647}{\em The Large N limit of superconformal field t} & J.M.Maldacena & 1 & \color{blue}16317 & \color{blue}6725 & \color{blue}22034\\ 
\hline2000 & \href{http://inspirehep.net/record/593382}{\em GEANT4--a simulation toolkit} & GEANT & 127 & \color{blue}13488 & \color{blue}6790 & 2170\\ 
2000 & \href{http://inspirehep.net/record/613135}{\em First year Wilkinson Microwave Anisotropy P} & WMAP & 17 & 9040 & 2990 & \color{blue}6507\\ 
\hline2010 & \href{http://inspirehep.net/record/1124337}{\em Observation of a new particle in the search} & ATLAS & 2932 & \color{blue}11666 & \color{blue}1652 & \color{blue}4074\\ 
\hline\end{tabular}
\end{center}
\caption{\em\label{tab:TopComparison} The top-cited, top-ranked, top-referred  paper written in each decennium
among those listed in \inspire (which is highly incomplete before 1960).\normalsize
}
\end{table}

The PaperRank in \eq{explicit}  splits papers into two qualitatively different categories:
\begin{itemize}

\item Sub-critical papers that get some attention and get forgotten: this happens when,
after a long time (tens of years), the sum over $g$ remains dominated by
the first generation.

\item Super-critical papers, that make history.
If the citation rate is high enough, it can sustain a `chain reaction', such that
late generations keep contributing significantly to the sum over generations.
At the same time, the original paper gets summarized in books and ceases to be cited directly.

\end{itemize}

\Fig{RankGen} shows, for a few notable papers,
how much different generations $g$ contribute to the sum in \eq{explicit}.
We see that about 10 generations contribute significantly for old top-ranked papers,
while the 1st generation provides the dominant contribution to  recent top-cited papers.\footnote{We computed
$R_p(\wp)$ analytically as function of $\wp$ for all papers $p$
using \eq{explicit} with the following  `pruning' algorithm.
To start, one finds all papers with no citations and eliminates them from the database,
after assigning their contributions to the $R_p$ of their references.
The process is iterated.
About 1000 iterations are needed to prune all the \inspire citation tree to nothing,
obtaining $R_p(\wp)$ as a power series in $\wp$.}

\begin{table}  \small\begin{center}
 \renewcommand{\arraystretch}{0.9}\setlength\tabcolsep{4pt}
\begin{tabular}{cllcc|ccc}
 & Title  & 1st author & $N_{\rm aut}$ & date  & $N_{\rm cit}$& $R_p$ & \color{blue}$\mathscr{R}_p$ \\  \hline 
1 & \href{http://inspirehep.net/record/1122534}{\em Black Holes: Complementarity or Firewall} & A.Almheiri & 4 & 2012 & 1129 & 230 & \color{blue}4712\\ 
2 & \href{http://inspirehep.net/record/1375326}{\em New Symmetries of QED} & D.Kapec & 3 & 2015 & 137 & 36 & \color{blue}1725\\ 
3 & \href{http://inspirehep.net/record/1467447}{\em Conformal symmetry and its breaking in t} & J.M.Maldacena & 3 & 2016 & 504 & 84 & \color{blue}1092\\ 
4 & \href{http://inspirehep.net/record/841876}{\em On the Origin of Gravity and the Laws of} & E.P.Verlinde & 1 & 2010 & 880 & 196 & \color{blue}1082\\ 
5 & \href{http://inspirehep.net/record/1452588}{\em Remarks on the Sachdev-Ye-Kitaev model} & J.M.Maldacena & 2 & 2016 & 860 & 173 & \color{blue}1059\\ 
6 & \href{http://inspirehep.net/record/1351297}{\em Topological insulators and superconducto} & X.L.Qi & 2 & 2011 & 1592 & 534 & \color{blue}1026\\ 
7 & \href{http://inspirehep.net/record/1236835}{\em Black holes and the butterfly effect} & S.H.Shenker & 2 & 2013 & 733 & 198 & \color{blue}939\\ 
8 & \href{http://inspirehep.net/record/846373}{\em Topological Insulators} & M.Zahid.Hasan & 2 & 2010 & 1892 & 788 & \color{blue}888\\ 
9 & \href{http://inspirehep.net/record/874812}{\em Shapiro Delay Measurement of A Two Solar} & P.Demorest & 5 & 2011 & 2411 & 396 & \color{blue}846\\ 
10 & \href{http://inspirehep.net/record/1242456}{\em Investigating the near-criticality of th} & D.Buttazzo & 7 & 2014 & 1086 & 127 & \color{blue}845
\end{tabular}
\caption{\em\label{tab:topreferred2010} Top-referred papers written after 2010 with less than 10 authors and more than 100 citations. }
\end{center}\normalsize\end{table}

\subsection{Top-referred (recent) papers}\label{sec:EarlyAlert}
As explained in the previous section, the PaperRank can single out some notable papers with few citations, provided that they are
old. However, when applied to recent papers (less than 10-20 years old), the PaperRank becomes highly correlated to the
number of (individual) citations, and therefore cannot perform better.

We thereby propose an early-alert indicator that recovers some information from the past.
First, we compute a rank among authors using all-time data.  
In particular, we adopt the AuthorRank $\mathscr{R}_A$ anticipated in \eq{Cmatrix} and better discussed in the next section.
Next, we use such rank to weight citations to papers, as in \eq{ARp}.
This means that we give more weight to citations from authors with higher $\mathscr{R}_A$,
implementing a sort of representative democracy.
We dub papers with top AuthorRank of papers $\mathscr{R}_p$  as `top-referred papers'.
\Tabl{topreferred0} shows the all-time list.

\Tabl{TopComparison} shows the top-cited, top-ranked and top-referred
papers published within each decennium, based on all subsequent citations.
For recent papers, top-cited and top-ranked tend to be dominated by
manuals of useful computer codes and by reviews.

We finally use $\mathscr{R}_p$ to find top-referred recent papers:
\tabl{topreferred2010} shows the top-referred papers published after year 2010
and with less than 10 authors (because our goal is to identify notable recent papers less known than discoveries made by
big experimental collaborations) and more than 100 citations.
Furthermore, we here removed self-citations, to avoid the list to be dominated by notable authors citing themselves.

\subsection{Paper metrics: correlations}
\begin{figure}[t]
$$\begin{array}{c}{\vspace{2cm}\includegraphics[width=0.25\textwidth]{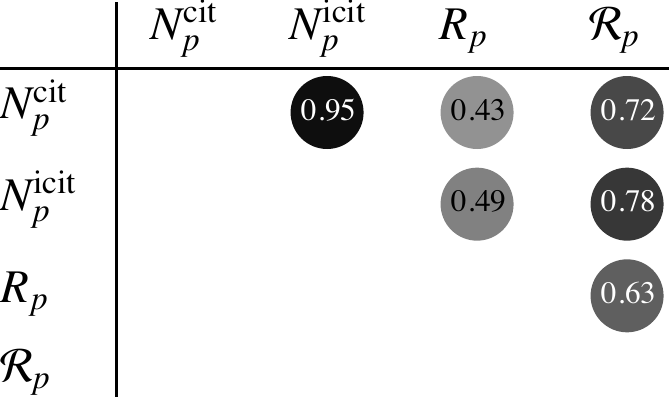}}\\ \,\vspace{6cm} \\ \end{array}\qquad \includegraphics[width=0.47\textwidth]{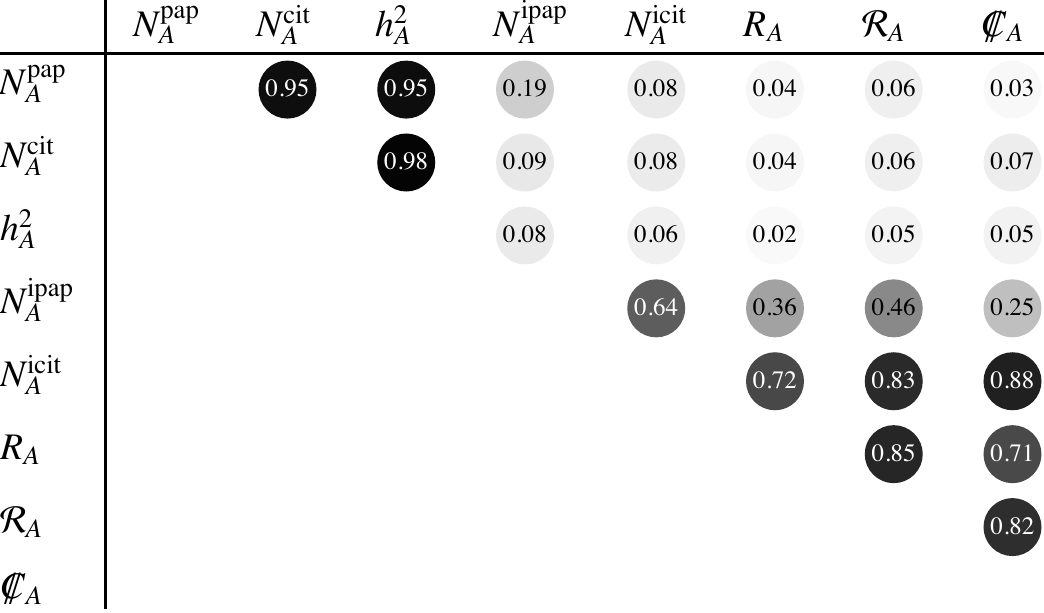}\vspace{-4.5cm}$$
\caption{\label{fig:Correlations}\em Left(right) table: correlations between indices of papers(authors). For papers the correlation is considered separately for the whole \inspire and for the eight main arXiv categories.}
\end{figure}

The left panel of \fig{Correlations} shows the correlations among the traditional counting of citations $N_{p}^{\text{cit}}$, the number of individual citations $N^{\rm icit}_{p}$, the PaperRank $R_{p}$ and the AuthorRank of papers $\mathscr{R}_{p}$, in the whole \inspire database. The number of individual citations is highly correlated with the number of citations.
Indeed, for papers, individual citations are just citations divided by the number of references of the citing papers so that uncorrelation is proportional to the variance around the average of the number of references per paper. The PaperRank and the AuthorRank are less correlated to the number of (individual) citations and to each other and represent fairly independent indices for ranking papers. 

\section{Ranking authors}\label{sec:AuthorRank}

\begin{table}[t] \small  \begin{center}
\renewcommand{\arraystretch}{0.9}\setlength\tabcolsep{1pt}
\begin{tabular}{clc}
  \multicolumn{3}{c}{Number of papers $\displaystyle\sum_{\rm papers} 1$} \\
  & InSpires name & All InSpires\\  \hline 
1& \inspiref{G.Eigen.1} & 2508\\
2& \inspiref{S.L.Wu.1} & 2396\\
3& \inspiref{Kazuhiko.Hara.1} & 2359\\
4& \inspiref{J.Brau.2} & 2310\\
5& \inspiref{A.Seiden.1} & 2293\\
6& \inspiref{R.Kass.1} & 2276\\
7& \inspiref{David.M.Strom.1} & 2270\\
8& \inspiref{A.Bodek.2} & 2268\\
9& \inspiref{W.T.Ford.1} & 2263\\
10& \inspiref{R.V.Kowalewski.1} & 2231
\end{tabular}~~~~
\begin{tabular}{clc}
  \multicolumn{3}{c}{Number of citations $\displaystyle\sum_{\rm papers} N_{\rm cit}$} \\
  & InSpires name & All InSpires\\  \hline 
1& \inspiref{R.V.Kowalewski.1} & 261176\\
2& \inspiref{G.D.Cowan.1} & 247656\\
3& \inspiref{Otmar.Biebel.1} & 236405\\
4& \inspiref{J.Huston.1} & 241391\\
5& \inspiref{Christoph.Grab.1} & 225873\\
6& \inspiref{Achim.Stahl.1} & 220954\\
7& \inspiref{K.Moenig.1} & 222876\\
8& \inspiref{S.L.Wu.1} & 225586\\
9& \inspiref{S.M.Spanier.1} & 215374\\
10& \inspiref{A.V.Gritsan.1} & 217487
\end{tabular}\\[2mm]
\begin{tabular}{clc}
  \multicolumn{3}{c}{$h$ index 
$\displaystyle\sum_{N_{\rm cit}\ge h} 1$}\\
  & InSpire name & All InSpires\\  \hline 
1& \inspiref{Kazuhiko.Hara.1} & 203\\
2& \inspiref{J.Huston.1} & 200\\
3& \inspiref{H.H.Williams.1} & 198\\
4& \inspiref{S.L.Wu.1} & 197\\
5& \inspiref{A.G.Clark.1} & 196\\
6& \inspiref{G.Eigen.1} & 196\\
7& \inspiref{P.K.Sinervo.1} & 195\\
8& \inspiref{M.J.Shochet.1} & 195\\
9& \inspiref{J.Proudfoot.1} & 195\\
10& \inspiref{S.M.Errede.1} & 195
\end{tabular}~~~~
\begin{tabular}{clc}
  \multicolumn{3}{c}{Average citations $\displaystyle\langle N_{\rm cit}\rangle\phantom{\sum_{1}}$}\\
  & InSpire name & All InSpire\\  \hline 
1& \inspiref{Y.Oohata.1} & 13488\\
2& \inspiref{N.Eiden.1} et al. & 7673\\
3& \inspiref{J.L.Chuma.1} et al. & 6744\\
4& \inspiref{S.Chowdhury.1} et al. & 6389\\
5& \inspiref{P.Schaffner.1} & 6104\\
6& \inspiref{S.B.Lugovsky.1} & 6043\\
7& \inspiref{K.S.Lugovsky.1} & 5844\\
8& \inspiref{C.Roumenin.1} et al. & 5746\\
9& \inspiref{H.Yusupov.2} & 5732\\
10& \inspiref{V.S.Lugovsky.1} & 5195
\end{tabular}
\end{center}
\caption{\em\label{tab:Npap} Authors listed according to traditional biblio-metric indices: 
total number of papers (top left), 
of citations (top right), $h$-index (bottom left), average number of citations per paper (bottom right). }
\end{table}

\begin{table}[p]\tiny\footnotesize \begin{center}
 \renewcommand{\arraystretch}{0.9}\setlength\tabcolsep{2pt}
\begin{tabular}{clc}
 & InSpire name & All  \\  \hline 
1& \inspiref{E.Witten.1} \Milner\DiracICTP & 3703\\
2& \inspiref{Steven.Weinberg.1} \Nobel\Milner & 2487\\
3& \inspiref{G.tHooft.1} \Nobel\Wolf & 1592\\
4& \inspiref{S.W.Hawking.1} \Milner\DiracIOP\Wolf & 1382\\
5& \inspiref{A.M.Polyakov.1} \Milner\DiracICTP & 986.2\\
6& \inspiref{F.A.Wilczek.1} \Nobel\DiracICTP\Sakurai & 902.6\\
7& \inspiref{J.M.Maldacena.1} \Milner\DiracICTP & 900.8\\
8& \inspiref{R.W.Jackiw.1} \DiracICTP & 888.3\\
9& \inspiref{J.S.Schwinger.1} \Nobel & 854.4\\
10& \inspiref{A.D.Linde.1} \Milner\DiracICTP & 833.4\\
11& \inspiref{T.Sjostrand.1} \Sakurai & 829.7\\
12& \inspiref{L.Susskind.1} \Sakurai & 799.6\\
13& \inspiref{S.L.Glashow.1} \Nobel & 788.0\\
14& \inspiref{H.M.Georgi.1} \DiracICTP\Sakurai & 775.0\\
15& \inspiref{N.Seiberg.1} \Milner\DiracICTP & 719.0\\
16& \inspiref{P.A.M.Dirac.1} \Nobel & 704.1\\
17& \inspiref{Sidney.R.Coleman.1} \DiracICTP & 675.3\\
18& \inspiref{David.J.Gross.1} \Nobel\DiracICTP\Sakurai & 665.9\\
19& \inspiref{C.Vafa.1} \Milner\DiracICTP & 643.0\\
20& \inspiref{S.J.Brodsky.1} \Sakurai & 626.4\\
21& \inspiref{J.R.Ellis.1} \DiracIOP & 617.5\\
22& \inspiref{K.G.Wilson.1} \Nobel\Wolf & 613.5\\
23& \inspiref{Abdus.Salam.1} \Nobel & 607.3\\
24& \inspiref{M.Luscher.1} \Planck & 589.9\\
25& \inspiref{J.D.Bjorken.1} \Wolf & 586.6\\
26& \inspiref{R.L.Jaffe.1} & 579.0\\
27& \inspiref{A.Strominger.1} \Milner\DiracICTP & 569.5\\
28& \inspiref{Ashoke.Sen.1} \Milner\DiracICTP & 567.5\\
29& \inspiref{G.Veneziano.1} \DiracICTP & 563.2\\
30& \inspiref{C.N.Yang.1} \Nobel & 548.7\\
31& \inspiref{J.Polchinski.1} \Milner\DiracICTP & 546.2\\
32& \inspiref{Rabindra.N.Mohapatra.1} & 540.2\\
33& \inspiref{S.Deser.1} & 526.5\\
34& \inspiref{Alexander.Vilenkin.1} & 525.5\\
35& \inspiref{R.P.Feynman.1} \Nobel & 524.9\\
36& \inspiref{L.Wolfenstein.1} \Sakurai & 524.6\\
37& \inspiref{Stephen.Louis.Adler.1} \DiracICTP\Sakurai & 520.4\\
38& \inspiref{M.Gell.Mann.1} \Nobel & 519.4\\
39& \inspiref{B.Zumino.1} \DiracICTP\Planck & 518.6\\
40& \inspiref{John.H.Schwarz.1} \Milner\DiracICTP & 518.4\\
41& \inspiref{M.A.Shifman.1} \DiracICTP\Sakurai & 505.9\\
42& \inspiref{E.V.Shuryak.1} & 493.8\\
43& \inspiref{M.B.Wise.1} \Sakurai & 485.9\\
44& \inspiref{G.W.Gibbons.1} & 484.8\\
45& \inspiref{J.D.Bekenstein.1} \Wolf & 484.1\\
46& \inspiref{A.A.Tseytlin.1} & 483.9\\
47& \inspiref{L.N.Lipatov.1} & 474.9\\
48& \inspiref{H.Leutwyler.1} & 471.9\\
49& \inspiref{T.D.Lee.1} \Nobel & 459.0\\
50& \inspiref{N.Isgur.1} \Sakurai & 456.9
\end{tabular}
\begin{tabular}{lc}
  InSpire name & After 2000  \\  \hline 
\inspiref{J.M.Maldacena.1} \Milner\DiracICTP & 306.1\\
\inspiref{S.D.Odintsov.1} & 276.1\\
\inspiref{E.Witten.1} \Milner\DiracICTP & 272.0\\
\inspiref{T.Padmanabhan.1} & 254.7\\
\inspiref{P.Z.Skands.1} & 254.4\\
\inspiref{T.Sjostrand.1} \Sakurai & 252.2\\
\inspiref{S.Nojiri.1} & 246.6\\
\inspiref{S.Mrenna.1} & 219.2\\
\inspiref{G.P.Salam.1} & 217.1\\
\inspiref{V.Springel.1} & 216.5\\
\inspiref{D.T.Son.1} & 212.5\\
\inspiref{Ashoke.Sen.1} \Milner\DiracICTP & 208.6\\
\inspiref{C.Vafa.1} \Milner\DiracICTP & 186.7\\
\inspiref{M.Cacciari.1} & 180.7\\
\inspiref{U.G.Meissner.1} & 180.6\\
\inspiref{P.Nason.1} & 176.7\\
\inspiref{Ernest.Ma.1} & 170.2\\
\inspiref{D.E.Kharzeev.1} & 160.3\\
\inspiref{S.Tsujikawa.1} & 158.9\\
\inspiref{A.Strominger.1} \Milner\DiracICTP & 156.3\\
\inspiref{Martin.Bojowald.1} & 152.6\\
\inspiref{V.A.Kostelecky.1} & 150.7\\
\inspiref{S.S.Gubser.1} & 148.4\\
\inspiref{S.Capozziello.1} & 145.4\\
\inspiref{J.Polchinski.1} \Milner\DiracICTP & 144.7\\
\inspiref{Alan.D.Martin.1} & 144.3\\
\inspiref{G.Amelino.Camelia.1} & 143.7\\
\inspiref{A.Loeb.1} & 143.1\\
\inspiref{D.W.Hooper.1} & 141.1\\
\inspiref{A.Ashtekar.1} & 141.1\\
\inspiref{F.Aharonian.1} & 139.6\\
\inspiref{N.Arkani.Hamed.1} \Milner & 138.3\\
\inspiref{Wayne.Hu.1} & 137.1\\
\inspiref{A.C.Fabian.1} & 136.8\\
\inspiref{Rong.Gen.Cai.1} & 136.5\\
\inspiref{G.R.Dvali.1} & 136.3\\
\inspiref{L.Susskind.1} \Sakurai & 136.1\\
\inspiref{E.V.Shuryak.1} & 135.8\\
\inspiref{U.W.Heinz.1} & 134.5\\
\inspiref{F.Karsch.1} & 131.7\\
\inspiref{N.Kidonakis.1} & 131.5\\
\inspiref{G.Soyez.1} & 131.1\\
\inspiref{A.D.Linde.1} \Milner\DiracICTP & 128.9\\
\inspiref{A.A.Tseytlin.1} & 128.3\\
\inspiref{A.Strumia.1} & 124.7\\
\inspiref{M.Visser.1} & 123.9\\
\inspiref{S.D.M.White.1} & 122.6\\
\inspiref{Nathan.J.Berkovits.1} & 122.2\\
\inspiref{S.Frixione.1} & 122.1\\
\inspiref{J.R.Ellis.1} \DiracIOP & 120.1
\end{tabular}
\begin{tabular}{lc}
  InSpire name & After 2010  \\  \hline 
\inspiref{S.D.Odintsov.1} & 82.9\\
\inspiref{J.M.Maldacena.1} \Milner\DiracICTP & 82.3\\
\inspiref{N.Kidonakis.1} & 77.1\\
\inspiref{E.Witten.1} \Milner\DiracICTP & 71.8\\
\inspiref{U.G.Meissner.1} & 63.3\\
\inspiref{A.Strominger.1} \Milner\DiracICTP & 62.2\\
\inspiref{C.de.Rham.1} & 62.1\\
\inspiref{S.Tsujikawa.1} & 60.4\\
\inspiref{S.Nojiri.1} & 59.5\\
\inspiref{S.Capozziello.1} & 57.2\\
\inspiref{D.Stanford.1} & 57.2\\
\inspiref{P.Z.Skands.1} & 57.1\\
\inspiref{L.Susskind.1} \Sakurai & 55.3\\
\inspiref{S.Sachdev.1} \DiracICTP & 50.0\\
\inspiref{M.Czakon.1} & 49.4\\
\inspiref{G.P.Salam.1} & 48.2\\
\inspiref{R.Venugopalan.1} & 47.7\\
\inspiref{M.Luscher.1} \Planck & 47.7\\
\inspiref{R.E.Kallosh.1} & 47.1\\
\inspiref{D.W.Hooper.1} & 46.4\\
\inspiref{R.B.Mann.1} & 45.6\\
\inspiref{S.Hod.1} & 45.4\\
\inspiref{F.Maltoni.1} & 45.4\\
\inspiref{R.C.Myers.1} & 45.2\\
\inspiref{A.D.Linde.1} \Milner\DiracICTP & 44.9\\
\inspiref{A.De.Felice.1} & 43.2\\
\inspiref{X.L.Qi.1} & 43.1\\
\inspiref{S.F.King.1} & 42.9\\
\inspiref{C.Bambi.1} & 42.5\\
\inspiref{B.Schenke.1} & 42.3\\
\inspiref{T.Padmanabhan.1} & 42.1\\
\inspiref{K.Hinterbichler.1} & 41.9\\
\inspiref{H.T.Janka.1} & 41.1\\
\inspiref{E.N.Saridakis.1} & 40.8\\
\inspiref{P.Nason.1} & 40.2\\
\inspiref{A.Mitov.1} & 40.1\\
\inspiref{Bing.Zhang.1} & 39.9\\
\inspiref{U.W.Heinz.1} & 39.3\\
\inspiref{Florian.R.A.Staub.1} & 39.3\\
\inspiref{A.Strumia.1} & 39.2\\
\inspiref{J.Rojo.1} & 39.1\\
\inspiref{T.Schwetz.1} & 39.0\\
\inspiref{A.J.Buras.1} & 39.0\\
\inspiref{A.Loeb.1} & 38.8\\
\inspiref{G.Soyez.1} & 38.7\\
\inspiref{E.Oset.1} & 38.1\\
\inspiref{V.Cardoso.1} & 37.8\\
\inspiref{E.P.Verlinde.1} & 37.2\\
\inspiref{S.S.Ostapchenko.1} & 36.9\\
\inspiref{P.Bozek.1} & 36.8
\end{tabular}
\end{center}
\caption{\em\label{tab:Icit} Authors sorted according to their number of individual citations $N_{A}^{\rm icit}$ for the whole \inspire database (left), from year 2000 (middle), and from year 2010 (right).
}
\end{table}

We start from the simplest and most naive  metrics:
in the top-left column of \tabl{Npap} we list the authors with most papers.
Within the \inspire database,
they are all experimentalists that  participate in  large collaborations with many co-authors.
The extreme case are the ATLAS and CMS collaborations
with  $\sim 10^3$ papers and $\sim 10^3$ authors.

In the top-right column of \tabl{Npap} we show the top-cited authors: again they
are experimentalists that  participate in large collaborations.
The citations of author $A$ are counted in the usual way:
summing the citations received by all papers that include $A$ as author, as in \eq{usual}.
The bottom-left column of \tabl{Npap} shows the authors with highest $h$ index,
and the bottom-right column the authors with the highest average number of citations per paper.

Next to each author we add symbols which show if they received the
Nobel (\Nobel), Dirac (\DiracICTP{} from ICTP and  \DiracIOP{} from IOP), Planck (\Planck),
Sakurai (\Sakurai), Wolf (\Wolf) and Milner (\Milner) prizes.
Small inaccuracies are possible, as medalists have been identified from names.
None of the top authors according to traditional metrics received any of them.

\smallskip

Clearly, all the indices shown in \tabl{Npap} ceased to be
relevant for experimentalists in view of the large number of co-authors.
This shows the need for an improved metrics that corrects for the inflation in the number of co-authors and allows at least a naive comparison of experimentalists with the rest of the community.

\subsection{Sharing among co-authors: fractional counting}
An improved metrics is obtained by attributing 
a fraction $p_{A}$ of any given paper to each author $A$, 
and imposing the sum rule $\sum p_A= 1$.
The fractions $p_A$ should tell how much each author contributed to the paper.
In the absence of this information, {\em we assume that each co-author contributed equally, so that $p_A= 1/N_p^{\rm aut}$}.\footnote{Weighting authors proportionally to their AuthorRank  
gives a non-linear system of equations, 
with singular solutions where a few notable authors collect all the weight of large collaborations.
Restricting to single-author papers would uniquely identify authors' contributions, but at the price of discarding most literature.}
This is called `fractional counting' in the bibliometric literature.

Taking into account this factor, the total number of {\em `individual papers'} of author $A$
is given by $\sum_{p\in A} 1/N_p^{\rm aut}$.
The same sharing among co-authors is applied to  citations.
The number of {\em `individual citations'} received by author $A$ is defined by summing over all its papers $p$ 
taking into account that citations are shared among co-authors, and weighted inversely to the number of references:
\beq\label{eq:IC}
N_A^{\rm icit}\equiv
\sum_{p\in A} \frac{N^{\rm icit}_p}{N^{\rm aut}_p}=
\sum_{p\in A}  \frac{1}{\hbox{(Number of authors of paper $p$)}}
\sum_{p'\to p}  \frac{1}{\hbox{(Number of references of $p'$)}}\,.\eeq
In the same way, we share the rank $R_p$ of each paper equally among its authors.
The rank $R_p$ of a paper approximates a physical quantity: how many times the paper is read.
The rank $R_A$ of an author inherits the same  meaning: it tells
the visibility of any author $A$, obtained by summing the visibility of its papers $p$ as
in \eq{ranks}, i.e.\footnote{For different methods see \refchange{\reffs{DBLP:journals/jasis/DingYFC09,DBLP:conf/icdm/ZhouOZG07}}{\cite{DBLP:journals/jasis/DingYFC09} and \cite{DBLP:conf/icdm/ZhouOZG07}}.
}
\beq\label{eq:AR}
R_A=
\hbox{(PaperRank of author $A$)} 
= \sum_{p\in A}  
\frac{\hbox{(PaperRank of paper $p$)}} {\hbox{(Number of authors of paper $p$)}}\,.\eeq
As discussed in \secc{rank}
we consider $\wp =0.99$.

\medskip

\begin{figure}[t]
$$
\includegraphics[width=0.55\textwidth]{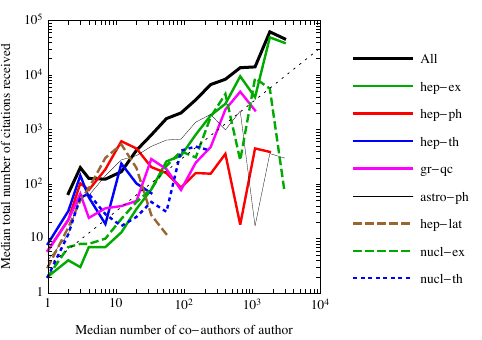}\qquad
\includegraphics[width=0.39\textwidth]{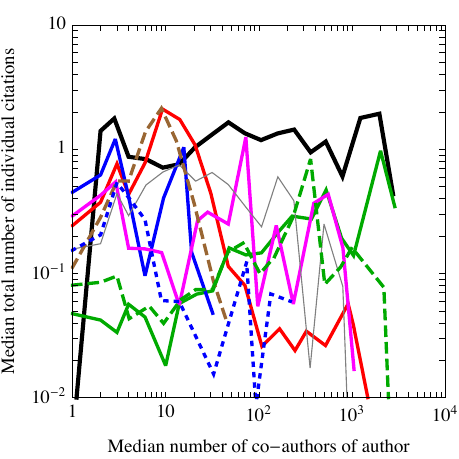}
$$
\caption{\em \label{fig:NautNcit} 
{\bf Left}: for each author in \inspire
we compute its mean number of co-authors 
(per paper, including the author itself) and 
the total number of citations of all its papers.
We do not show the scatter plot, only its median,
separately for each arXiv category.
In both cases the total number of citations grows roughly linearly with the number of authors.
{\bf Right}: As in the middle panel, but showing on the vertical axis the number of individual citations,
which roughly does not depend on the number of authors.
}
\end{figure}


\subsection{Fractional counting and collaborations}\label{sec:FC}
We share the number of citations received by a paper among its $N_{\rm aut}$
authors as $1/N_{\rm aut}$.
Some agencies adopt various less steep functions of $N_{\rm aut}$.
This raises the question: how does the total number of citations received scale,
on average, with the number of co-authors?\footnote{We thank Paolo Rossi for raising this issue.}
\Fig{NautNcit} (described in more details in the caption)
provides the answer: the total number of citations received by an author
grows roughly linearly with the mean number of co-authors.
The right panel of \fig{NautNcit} shows that, instead, the  total number of individual citations
is scale-invariant, namely roughy independent of the number of  co-authors.
This means that individual citations do not
reward nor penalise big collaborations,
while citations reward big collaborations.



\begin{table}[t]
\tiny\footnotesize\renewcommand{\arraystretch}{0.9}\setlength\tabcolsep{2pt}
\begin{center}
\begin{tabular}{clc}
&InSpire name & All InSpire\\  \hline 
1& \inspiref{Steven.Weinberg.1}  \Nobel\Milner & 235825\\
2& \inspiref{J.S.Schwinger.1}  \Nobel & 231194\\
3& \inspiref{R.P.Feynman.1}  \Nobel & 162394\\
4& \inspiref{M.Gell.Mann.1}  \Nobel & 156387\\
5& \inspiref{C.N.Yang.1}  \Nobel & 114656\\
6& \inspiref{P.A.M.Dirac.1}  \Nobel & 111083\\
7& \inspiref{Abdus.Salam.1}  \Nobel & 104766\\
8& \inspiref{E.Witten.1}  \Milner\DiracICTP & 102753\\
9& \inspiref{H.A.Bethe.1}  \Nobel & 102195\\
10& \inspiref{G.tHooft.1}  \Nobel\Wolf & 99645\\
11& \inspiref{E.P.Wigner.1}  \Nobel\Planck & 90966\\
12& \inspiref{T.D.Lee.1}  \Nobel & 86699\\
13& \inspiref{W.Heisenberg.1}  \Nobel\Planck & 81238\\
14& \inspiref{Stephen.Louis.Adler.1}  \DiracICTP\Sakurai & 76687\\
15& \inspiref{Yoichiro.Nambu.1}  \Nobel\DiracICTP\Planck\Sakurai\Wolf & 74084\\
16& \inspiref{K.G.Wilson.1}  \Nobel\Wolf & 69320\\
17& \inspiref{S.L.Glashow.1}  \Nobel & 69308\\
18& \inspiref{F.J.Dyson.1}  \Planck\Wolf & 68021\\
19& \inspiref{A.M.Polyakov.1}  \Milner\DiracICTP & 67138\\
20& \inspiref{J.D.Bjorken.1}  \Wolf & 63817\\
21& \inspiref{S.W.Hawking.1}  \Milner\DiracIOP\Wolf & 62039\\
22& \inspiref{B.Zumino.1}  \DiracICTP\Planck & 58680\\
23& \inspiref{S.Mandelstam.1}  \DiracICTP & 57268\\
24& \inspiref{G.Breit.2}  & 52537\\
25& \inspiref{V.F.Weisskopf.1}  & 52355\\
26& \inspiref{Enrico.Fermi.1}  \Nobel\Planck & 52039\\
27& \inspiref{J.R.Oppenheimer.1}  & 51472\\
28& \inspiref{David.J.Gross.1}  \Nobel\DiracICTP\Sakurai & 50937\\
29& \inspiref{Peter.W.Higgs.1}  \Nobel\DiracIOP\Sakurai\Wolf & 50761\\
30& \inspiref{J.A.Wheeler.1}  \Wolf & 50238
\end{tabular}
\begin{tabular}{lc}
InSpire name & After 2000\\  \hline 
\inspiref{E.Witten.1}  \Milner\DiracICTP & 3670\\
\inspiref{J.M.Maldacena.1}  \Milner\DiracICTP & 3551\\
\inspiref{S.D.Odintsov.1}  & 3027\\
\inspiref{T.Sjostrand.1}  \Sakurai & 3016\\
\inspiref{D.T.Son.1}  & 2958\\
\inspiref{T.Padmanabhan.1}  & 2824\\
\inspiref{P.Z.Skands.1}  & 2804\\
\inspiref{S.Nojiri.1}  & 2783\\
\inspiref{V.Springel.1}  & 2701\\
\inspiref{A.Loeb.1}  & 2594\\
\inspiref{C.Vafa.1}  \Milner\DiracICTP & 2497\\
\inspiref{Ashoke.Sen.1}  \Milner\DiracICTP & 2468\\
\inspiref{G.P.Salam.1}  & 2453\\
\inspiref{Martin.Bojowald.1}  & 2373\\
\inspiref{S.Mrenna.1}  & 2352\\
\inspiref{Ernest.Ma.1}  & 2297\\
\inspiref{A.C.Fabian.1}  & 2285\\
\inspiref{F.Aharonian.1}  & 2189\\
\inspiref{Wayne.Hu.1}  & 2162\\
\inspiref{P.Nason.1}  & 2132\\
\inspiref{E.V.Shuryak.1}  & 2121\\
\inspiref{G.R.Dvali.1}  & 2100\\
\inspiref{U.G.Meissner.1}  & 2081\\
\inspiref{S.S.Gubser.1}  & 2003\\
\inspiref{D.E.Kharzeev.1}  & 1982\\
\inspiref{A.A.Tseytlin.1}  & 1946\\
\inspiref{M.Cacciari.1}  & 1937\\
\inspiref{M.Zaldarriaga.1}  & 1911\\
\inspiref{U.W.Heinz.1}  & 1876\\
\inspiref{Joseph.I.Silk.1}  & 1873
\end{tabular}
\begin{tabular}{lc}
InSpire name & After 2010\\  \hline 
\inspiref{N.Kidonakis.1}  & 822.3\\
\inspiref{E.Witten.1}  \Milner\DiracICTP & 794.2\\
\inspiref{J.M.Maldacena.1}  \Milner\DiracICTP & 786.5\\
\inspiref{U.G.Meissner.1}  & 785.3\\
\inspiref{S.Sachdev.1}  \DiracICTP & 776.9\\
\inspiref{S.Hod.1}  & 752.2\\
\inspiref{D.Stanford.1}  & 715.1\\
\inspiref{X.L.Qi.1}  & 713.0\\
\inspiref{S.D.Odintsov.1}  & 704.1\\
\inspiref{Muhammad.Sharif.1}  & 701.7\\
\inspiref{A.Loeb.1}  & 700.3\\
\inspiref{S.Capozziello.1}  & 674.2\\
\inspiref{C.de.Rham.1}  & 657.4\\
\inspiref{L.Susskind.1}  \Sakurai & 650.0\\
\inspiref{S.Tsujikawa.1}  & 588.1\\
\inspiref{Bing.Zhang.1}  & 577.6\\
\inspiref{D.W.Hooper.1}  & 573.5\\
\inspiref{C.Bambi.1}  & 564.9\\
\inspiref{R.B.Mann.1}  & 564.0\\
\inspiref{A.Strominger.1}  \Milner\DiracICTP & 548.2\\
\inspiref{E.Oset.1}  & 534.9\\
\inspiref{P.Z.Skands.1}  & 526.5\\
\inspiref{Xiao.Gang.Wen.1}  & 520.4\\
\inspiref{S.Nojiri.1}  & 514.0\\
\inspiref{H.T.Fortune.1}  & 512.1\\
\inspiref{S.F.King.1}  & 503.2\\
\inspiref{J.R.Ellis.1}  \DiracIOP & 500.6\\
\inspiref{C.L.Kane.1}  \Milner\DiracICTP & 500.0\\
\inspiref{R.Myrzakulov.1}  & 497.8\\
\inspiref{A.Vishwanath.1}  & 494.5
\end{tabular}
\end{center}
\caption{\em\label{tab:PR} Authors sorted according to the PaperRank of authors, $R_A = \sum_{\rm papers} R_p/N_{\rm aut}$ for the whole \inspire database (left), from year 2000 (middle), and from year 2010 (right).}
\end{table}

\subsection{PaperRank of authors: results}
We now apply the improved metrics described in the previous section to the \inspire database.

The left column of \tabl{Icit} lists the authors with the highest number of individual citations.
Two factors  differentiate citations from  individual citations.
First, dividing by the number of references (first factor in the denominator in \eq{IC})
counter-acts the inflation in the total number of citations.
This factors mildly penalises authors working in sectors (such as hep-ph) where papers have a
larger average number of references.
Second, dividing by the number of authors (second factor at the denominator in \eq{IC}) has a large impact:
members of huge collaborations no longer make the top positions of the list,
which becomes dominated by theorists.
As discussed in \secc{FC}, this happens 
because working in large collaborations does
not allow to recognise individual merit --- not because working in large collaborations decreases the average merit.
Lists of bottom authors would similarly be dominated by theorists.

\medskip

The left column of \tabl{PR} shows the top-ranked authors  in the \inspire database.
The PaperRank of authors identifies some older notable authors who received the prizes
plotted in front of their name, despite having
less citations than modern authors, given the increase in the rate of papers and of citations.

\medskip

Anyhow, the main interest of our study is not re-discovering Feynman.
We want to see if our metrics do a better job than just citation counts in identifying modern authors with a high impact.
To achieve this, we set a lower cut-off on the publication year.
We restrict the list to papers published `From 2000' (middle column of \tabl{PR})
and `From 2010' (right column of \tabl{PR}).

While switching from citations to individual citations is an obvious improvement,
we find that the rank does not improve over individual citations (they are strongly correlated)
when restricting to recent papers.
As already discussed for papers, about 10-20 years are needed before that the PaperRank becomes a better metrics.
On shorter time-scales, the rank and the number of individual citations are strongly correlated, and no significant differences arise;
a few authors have a rank significantly higher than their number of individual citations
often because they happen to be cited by reviews which fast received a large number of citations.
Restricting the sums to the $N$-th best papers of each author has little effect.

\medskip

\begin{table}[t]
\tiny\footnotesize\renewcommand{\arraystretch}{0.9}\setlength\tabcolsep{2pt}
\begin{center}
\begin{tabular}{clc}
 & InSpire name & All  \\  \hline 
1& \inspiref{P.A.M.Dirac.1} \Nobel & 292479\\
2& \inspiref{Albert.Einstein.1} \Nobel\Planck & 229323\\
3& \inspiref{E.Witten.1} \Milner\DiracICTP & 166374\\
4& \inspiref{Steven.Weinberg.1} \Nobel\Milner & 147661\\
5& \inspiref{G.tHooft.1} \Nobel\Wolf & 109341\\
6& \inspiref{J.S.Schwinger.1} \Nobel & 98088\\
7& \inspiref{Max.Born.1} \Nobel\Planck & 81451\\
8& \inspiref{S.W.Hawking.1} \Milner\DiracIOP\Wolf & 79424\\
9& \inspiref{A.M.Polyakov.1} \Milner\DiracICTP & 69293\\
10& \inspiref{R.P.Feynman.1} \Nobel & 68094\\
11& \inspiref{M.Gell.Mann.1} \Nobel & 66393\\
12& \inspiref{C.N.Yang.1} \Nobel & 65056\\
13& \inspiref{Enrico.Fermi.1} \Nobel\Planck & 64907\\
14& \inspiref{K.G.Wilson.1} \Nobel\Wolf & 54851\\
15& \inspiref{H.A.Bethe.1} \Nobel & 53697\\
16& \inspiref{T.D.Lee.1} \Nobel & 50413\\
17& \inspiref{L.Susskind.1} \Sakurai & 46873\\
18& \inspiref{Abdus.Salam.1} \Nobel & 46278\\
19& \inspiref{S.L.Glashow.1} \Nobel & 45184\\
20& \inspiref{R.W.Jackiw.1} \DiracICTP & 44070\\
21& \inspiref{F.A.Wilczek.1} \Nobel\DiracICTP\Sakurai & 43153\\
22& \inspiref{H.M.Georgi.1} \DiracICTP\Sakurai & 40178\\
23& \inspiref{Stephen.Louis.Adler.1} \DiracICTP\Sakurai & 39928\\
24& \inspiref{E.P.Wigner.1} \Nobel\Planck & 39752\\
25& \inspiref{Sidney.R.Coleman.1} \DiracICTP & 39247\\
26& \inspiref{David.J.Gross.1} \Nobel\DiracICTP\Sakurai & 38777\\
27& \inspiref{A.D.Linde.1} \Milner\DiracICTP & 38338\\
28& \inspiref{J.D.Bjorken.1} \Wolf & 38272\\
29& \inspiref{W.Heisenberg.1} \Nobel\Planck & 36412\\
30& \inspiref{S.Mandelstam.1} \DiracICTP & 35261
\end{tabular}
\begin{tabular}{lc}
  InSpire name & After 2000  \\  \hline 
\inspiref{E.Witten.1} \Milner\DiracICTP & 18748\\
\inspiref{T.Sjostrand.1} \Sakurai & 16669\\
\inspiref{J.M.Maldacena.1} \Milner\DiracICTP & 16406\\
\inspiref{V.Springel.1} & 15267\\
\inspiref{P.Z.Skands.1} & 13705\\
\inspiref{S.Mrenna.1} & 11931\\
\inspiref{C.Vafa.1} \Milner\DiracICTP & 11840\\
\inspiref{G.P.Salam.1} & 11091\\
\inspiref{D.T.Son.1} & 10511\\
\inspiref{P.Nason.1} & 10405\\
\inspiref{J.A.M.Vermaseren.1} & 9195\\
\inspiref{Alan.D.Martin.1} & 8994\\
\inspiref{S.D.M.White.1} & 8761\\
\inspiref{B.R.Webber.1} & 8673\\
\inspiref{A.Loeb.1} & 8519\\
\inspiref{Ashoke.Sen.1} \Milner\DiracICTP & 8427\\
\inspiref{L.E.Hernquist.1} & 8331\\
\inspiref{J.Polchinski.1} \Milner\DiracICTP & 8193\\
\inspiref{M.Cacciari.1} & 8146\\
\inspiref{S.Frixione.1} & 8091\\
\inspiref{L.Susskind.1} \Sakurai & 8030\\
\inspiref{A.C.Fabian.1} & 7809\\
\inspiref{Wayne.Hu.1} & 7797\\
\inspiref{A.Strominger.1} \Milner\DiracICTP & 7750\\
\inspiref{M.Luscher.1} \Planck & 7710\\
\inspiref{Nathan.J.Berkovits.1} & 7679\\
\inspiref{Xiao.Gang.Wen.1} & 7623\\
\inspiref{U.G.Meissner.1} & 7613\\
\inspiref{A.Y.Kitaev.1} \Milner\DiracICTP & 7472\\
\inspiref{F.Aharonian.1} & 7438
\end{tabular}
\begin{tabular}{lc}
  InSpire name & After 2010  \\  \hline 
\inspiref{N.Kidonakis.1} & 7248\\
\inspiref{E.Witten.1} \Milner\DiracICTP & 5178\\
\inspiref{J.M.Maldacena.1} \Milner\DiracICTP & 4799\\
\inspiref{P.Z.Skands.1} & 4335\\
\inspiref{A.A.Abdo.1} & 4294\\
\inspiref{M.Czakon.1} & 4160\\
\inspiref{D.Stanford.1} & 4160\\
\inspiref{X.L.Qi.1} & 3970\\
\inspiref{M.Luscher.1} \Planck & 3929\\
\inspiref{G.Harry.1} & 3666\\
\inspiref{L.Susskind.1} \Sakurai & 3322\\
\inspiref{Alexander.Romanenko.1} & 3318\\
\inspiref{Xiao.Gang.Wen.1} & 3249\\
\inspiref{S.Sachdev.1} \DiracICTP & 3224\\
\inspiref{A.Strominger.1} \Milner\DiracICTP & 3202\\
\inspiref{H.T.Janka.1} & 3108\\
\inspiref{P.Nason.1} & 3089\\
\inspiref{J.Rojo.1} & 3083\\
\inspiref{A.Mitov.1} & 3077\\
\inspiref{G.P.Salam.1} & 2934\\
\inspiref{C.L.Kane.1} \Milner\DiracICTP & 2831\\
\inspiref{F.Maltoni.1} & 2775\\
\inspiref{A.Vishwanath.1} & 2739\\
\inspiref{U.G.Meissner.1} & 2702\\
\inspiref{J.Polchinski.1} \Milner\DiracICTP & 2649\\
\inspiref{Patrick.Huber.1} & 2606\\
\inspiref{C.de.Rham.1} & 2577\\
\inspiref{A.Loeb.1} & 2521\\
\inspiref{E.Berger.1} & 2492\\
\inspiref{S.Forte.2} & 2463
\end{tabular}
\end{center}
\caption{\em\label{tab:AR} Authors sorted according to their Author Rank $\mathscr{R}_A$ for the whole \inspire database (left), from year 2000 (middle), and from year 2010 (right).}
\end{table}

\subsection{Author Rank}\label{sec:authorrank}
As outlined in the introduction, the citation matrix between authors (properly normalized) $C_{A'\to A}$ defined in \eq{Cmatrix} allows to define an AuthorRank as
\beq \mathscr{R}_A = \wp \sum_{A'} \mathscr{R}_{A'} C_{A'\to A}   +\alpha(1- \wp)\,,\eeq
where the second term gives a constant weight 
to each author, independently from the number and quality of its papers.
{The network of citations among authors avoids time-directness, up to time-scales comparable to the scientific ages of authors (which is enough for some goals, such as studying how senior authors evaluate the work of younger authors).}
While formally analogous to the ranking of papers, this ranking of authors is not a model of a physical process, because one reads papers, not authors.
The graph corresponding to the matrix $C_{A'\to A}$ contains cycles and also loops on the same node (self-citations):
here we use $\wp=0.9$, such that self-citations cannot boost the AuthorRank by more than one
order of magnitude.

The left column of \tabl{AR} shows the all-time AuthorRank.
We see the emergence of old authors such as Einstein,
which were absent from previous top-rankings because
poorly covered and cited in the too recent {\sc InSpire} database.
Of course, the incompleteness of \inspire before $\sim 1960$ makes results about older authors semi-quantitative.

This issue is avoided in the other columns of \tabl{AR}, 
where we show the AuthorRank recomputed by restricting to recent papers only.

\begin{table}[t]
\tiny \footnotesize\renewcommand{\arraystretch}{0.9}\setlength\tabcolsep{2pt}
\begin{center}
\begin{tabular}{clc}
 & InSpire name & All  \\  \hline 
1& \inspiref{E.Witten.1}  \Milner\DiracICTP & 3428\\
2& \inspiref{Steven.Weinberg.1}  \Nobel\Milner & 2261\\
3& \inspiref{G.tHooft.1}  \Nobel\Wolf & 1366\\
4& \inspiref{S.W.Hawking.1}  \Milner\DiracIOP\Wolf & 1217\\
5& \inspiref{A.M.Polyakov.1}  \Milner\DiracICTP & 923.5\\
6& \inspiref{J.M.Maldacena.1}  \Milner\DiracICTP & 824.3\\
7& \inspiref{J.S.Schwinger.1}  \Nobel & 739.7\\
8& \inspiref{T.Sjostrand.1}  \Sakurai & 736.1\\
9& \inspiref{F.A.Wilczek.1}  \Nobel\DiracICTP\Sakurai & 673.7\\
10& \inspiref{L.Susskind.1}  \Sakurai & 658.6\\
11& \inspiref{R.W.Jackiw.1}  \DiracICTP & 655.8\\
12& \inspiref{A.D.Linde.1}  \Milner\DiracICTP & 652.6\\
13& \inspiref{S.L.Glashow.1}  \Nobel & 636.8\\
14& \inspiref{N.Seiberg.1}  \Milner\DiracICTP & 627.8\\
15& \inspiref{P.A.M.Dirac.1}  \Nobel & 624.4\\
16& \inspiref{Sidney.R.Coleman.1}  \DiracICTP & 622.9\\
17& \inspiref{H.M.Georgi.1}  \DiracICTP\Sakurai & 614.4\\
18& \inspiref{K.G.Wilson.1}  \Nobel\Wolf & 552.5\\
19& \inspiref{David.J.Gross.1}  \Nobel\DiracICTP\Sakurai & 543.2\\
20& \inspiref{C.Vafa.1}  \Milner\DiracICTP & 521.4\\
21& \inspiref{M.Luscher.1}  \Planck & 508.0\\
22& \inspiref{R.P.Feynman.1}  \Nobel & 459.9\\
23& \inspiref{A.Strominger.1}  \Milner\DiracICTP & 450.9\\
24& \inspiref{R.L.Jaffe.1}  & 441.2\\
25& \inspiref{J.Polchinski.1}  \Milner\DiracICTP & 439.7\\
26& \inspiref{J.D.Bjorken.1}  \Wolf & 422.5\\
27& \inspiref{M.Gell.Mann.1}  \Nobel & 420.0\\
28& \inspiref{C.N.Yang.1}  \Nobel & 415.9\\
29& \inspiref{B.Zumino.1}  \DiracICTP\Planck & 412.8\\
30& \inspiref{Abdus.Salam.1}  \Nobel & 410.0
\end{tabular}
\begin{tabular}{lc}
  InSpire name & After 2000  \\  \hline 
\inspiref{J.M.Maldacena.1}  \Milner\DiracICTP & 253.5\\
\inspiref{T.Sjostrand.1}  \Sakurai & 228.8\\
\inspiref{P.Z.Skands.1}  & 223.0\\
\inspiref{S.Mrenna.1}  & 207.7\\
\inspiref{E.Witten.1}  \Milner\DiracICTP & 197.0\\
\inspiref{S.D.Odintsov.1}  & 192.1\\
\inspiref{G.P.Salam.1}  & 181.2\\
\inspiref{S.Nojiri.1}  & 173.1\\
\inspiref{D.T.Son.1}  & 162.8\\
\inspiref{V.Springel.1}  & 157.4\\
\inspiref{T.Padmanabhan.1}  & 156.6\\
\inspiref{M.Cacciari.1}  & 151.7\\
\inspiref{P.Nason.1}  & 148.5\\
\inspiref{C.Vafa.1}  \Milner\DiracICTP & 134.7\\
\inspiref{Ashoke.Sen.1}  \Milner\DiracICTP & 125.7\\
\inspiref{N.Arkani.Hamed.1}  \Milner & 119.6\\
\inspiref{A.Strominger.1}  \Milner\DiracICTP & 119.5\\
\inspiref{V.A.Kostelecky.1}  & 113.1\\
\inspiref{J.Polchinski.1}  \Milner\DiracICTP & 108.0\\
\inspiref{G.Soyez.1}  & 99.7\\
\inspiref{D.E.Kharzeev.1}  & 97.6\\
\inspiref{P.Horava.1}  & 97.3\\
\inspiref{S.S.Gubser.1}  & 96.5\\
\inspiref{M.Luscher.1}  \Planck & 94.6\\
\inspiref{S.Tsujikawa.1}  & 93.8\\
\inspiref{L.Susskind.1}  \Sakurai & 93.7\\
\inspiref{S.Frixione.1}  & 92.7\\
\inspiref{N.Seiberg.1}  \Milner\DiracICTP & 90.3\\
\inspiref{A.Ashtekar.1}  & 89.4\\
\inspiref{M.A.Stephanov.1}  & 87.5
\end{tabular}
\begin{tabular}{lc}
  InSpire name & After 2010  \\  \hline 
\inspiref{J.M.Maldacena.1}  \Milner\DiracICTP & 68.0\\
\inspiref{D.Stanford.1}  & 50.2\\
\inspiref{S.D.Odintsov.1}  & 49.0\\
\inspiref{C.de.Rham.1}  & 48.7\\
\inspiref{E.Witten.1}  \Milner\DiracICTP & 48.2\\
\inspiref{A.Strominger.1}  \Milner\DiracICTP & 47.9\\
\inspiref{N.Kidonakis.1}  & 47.2\\
\inspiref{P.Z.Skands.1}  & 44.7\\
\inspiref{M.Luscher.1}  \Planck & 39.6\\
\inspiref{G.P.Salam.1}  & 38.5\\
\inspiref{M.Czakon.1}  & 38.2\\
\inspiref{L.Susskind.1}  \Sakurai & 36.9\\
\inspiref{F.Maltoni.1}  & 34.6\\
\inspiref{S.Nojiri.1}  & 34.0\\
\inspiref{E.P.Verlinde.1}  & 33.5\\
\inspiref{R.C.Myers.1}  & 33.1\\
\inspiref{S.Tsujikawa.1}  & 32.2\\
\inspiref{A.Mitov.1}  & 32.0\\
\inspiref{G.Soyez.1}  & 30.8\\
\inspiref{P.Nason.1}  & 30.6\\
\inspiref{M.Cacciari.1}  & 29.7\\
\inspiref{A.D.Linde.1}  \Milner\DiracICTP & 28.4\\
\inspiref{A.A.Abdo.1}  & 28.2\\
\inspiref{Olivier.Mattelaer.1}  & 28.0\\
\inspiref{A.De.Felice.1}  & 27.9\\
\inspiref{X.L.Qi.1}  & 26.8\\
\inspiref{D.Simmons.Duffin.1}  & 26.7\\
\inspiref{K.Hinterbichler.1}  & 26.4\\
\inspiref{R.Venugopalan.1}  & 26.4\\
\inspiref{N.Seiberg.1}  \Milner\DiracICTP & 26.2
\end{tabular}
\end{center}
\caption{\em\label{tab:CCA} 
Authors sorted according to their Citation-coin $\CC_A$ of \eq{CCdef} for the whole \inspire database (left), from year 2000 (middle), and from year 2010 (right)..}
\end{table}

\subsection{Removing self-citations and citation `cartels': the Citation-coin}\label{sec:self}\label{sec:cartels}
One of the unsatisfactory aspects of previous metrics is the effect of self-citations.

On short time-scales  the PaperRank $R$ and the number of (individual) citations are strongly correlated, so they can be
similarly inflated trough self-citations.
Only on longer time-scales $R$ becomes a more robust measure: 
citations from paper $p'$ are weighted by its rank $R_{p'}$, giving relatively less weight to `below average' papers that sometimes contain many self-citations.
Still, many below average papers can sum up to a significant total rank.

One can optionally count citations from published papers only, ignoring citations from unpublished papers.
However this choice discards information on good unpublished papers
(some well respected authors do not publish some of their papers).

Removing all self-citations is, by itself, an arbitrary choice.
Furthermore it can be implemented in different ways, for example removing citations from co-authors.
Removing only citations of an author to itself amounts to set to zero the diagonal
elements of the citation matrix $N^{\rm icit}_{A'\to A}$,
reducing its $N^2$ entries to $N(N-1)$.

This does not protect from `citation cartels'.
A second step in this direction consists in removing citations exchanges $A\to A'$ and $A'\to A$ between all pairs of authors $A,A'$.
This amounts to subtract the symmetric part of the $N^{\rm icit}_{A\to A'}$ matrix, reducing its entries to $N(N-1)/2$.

A third step is removing citations exchanges $A\to A'$, $A'\to A''$ and $A''\to A$ among triplets of authors $A,A',A''$.  
A fourth step is removing all quadruplets etc.

A combinatorial computation shows that, 
after removing all possible `cartels', only $N-1$ entries remain.
They can be described by $N$ numbers $\CC_A$ that sum up to 0.
The meaning of $\CC_A$ can be intuitively understood by viewing $N^{\rm icit}_{A'\to A}$ as the total amount `paid' by $A'$ to $A$,
and  $N^{\rm icit}$ as a matrix of transactions.
Then the physical quantity unaffected by cartels 
is the net amount  `owned' by each author $A$:
$ \CC_A = \sum_{A'} (N^{\rm icit}_{A'\to A}-N^{\rm icit}_{A\to A'}).$ 
Citations are treated like money:
subtracting all possible citation
`cartels' is equivalent to count citations received as positive, and citations given as negative.
In doing this we proceed as described above: we actually count individual citations (`icit'):
shared between co-authors, and divided by the number of references of each paper.
Then the price paid is the total number of papers written, independently of their number of references.
The Citation-coin of authors, $\CC_A$, can be written in terms of the Citation-coin of their papers, $\CC_p$:
\beq \label{eq:CCdef}
\CC_A = \sum_{p\in A} \frac{\CC_p}{N_p^{\rm aut}}\,,\qquad
\CC_p = N_p^{\rm icit}-f( t_{\rm end}-t_p )\eeq
with $f( t_{\rm end}-t_p )=1$. A paper has $\CC_p<0$ when it receives a below-average number of individual citations.
This is the case for all recent papers, suggesting the introduction of a factor $f( t_{\rm end}-t_p )\le 1$.
All metrics penalise recent papers and authors 
(namely those active a few years before $t_{\rm end}$, the end-date of the database),
because citations accumulate over time.
This penalisation is stronger in the Citation-coin than in citation counting,
if a paper $p$ is paid at the moment of its appearance $t_p$.
This boundary effect can be compensated by redefining the Citation-coin
in such a way that papers are `paid in instalments',
with the same time-scale over which citations accumulate.
One can conveniently choose $f(t) \approx 1- e^{-t/\tau}$, where $\tau \approx 7\,{\rm yr}$
is presently the  time-scale over which citations are received by an average paper in the \inspire{} database.\footnote{This number is obtained considering the whole \inspire and it varies by a factor of around two across the main arXiv categories. As for fractional counting, in the case one wants to fine-tune comparisons within a given sub-field, this number can be tuned to the behavior of that sub-field.}
In this way $f(t) \simeq 1$ for old papers, $f(t) < 1$ for young papers,
$f(t)\simeq0$ for very recent papers.
As a consequence recent authors do not necessarily have a negative Citation-coin,
and the lack of knowledge of future citations does not penalise their $\CC$ more than their number 
of citations.\footnote{Furthermore, one can define a $\CC^+$ metric which discards `negative' papers with $\CC_p <0$ and sums the contributions of `positive' papers.}

\medskip

\Tabl{CCA} lists authors according to their Citation-coin $\CC_A$.
Authors that scored highly in previous ranks by writing many papers with low impact have now disappeared, and some of them actually got a negative $\CC_A$.

It is interesting to compare the Citation-coin to traditional metrics. Unlike the number of (individual) citations, and unlike the number of papers,
the $\CC$ does not reward authors that publish many low impact papers. Like the $h$-index, the $\CC$ rewards both quality and quantity,
but in a different way.
The $h$-index puts a threshold on the number of citations: as a result
papers below don't contribute, and all papers above count equally.
On the other hand, excellent papers contribute significantly to the $\CC$,
while below-average papers contribute negatively. In this respect, the Citation-coin can be considered an improvement of the $h$-index that is both more indicative and less correlated (see \secc{corr}) with citation counting.
The  $\CC$ also differs from $\langle N_{\rm cit}\rangle$, 
the number of (possibly individual) citations averaged over all papers 
of the author under consideration.
Unlike the $\CC$, $\langle N_{\rm cit}\rangle$ only rewards quality: 
an author can maximise it by  writing only very few excellent papers. At the practical level,
the $\CC$ identifies physicists considered most esteemed (see \tabl{CCA}), while
traditional metrics fail (see \tabl{Npap}).

\smallskip

Finally, from the database we can extract detailed reports about the metrics  of each author:
$N_{\rm pap}$, $N_{\rm cit}$, $h$, $N_{\rm ipap}$, $N_{\rm icit}$, $R$, $\mathscr{R}$, $\CC$,
their time evolution, the scientific age,
the percentage of given and received self-citations, 
the topics studied, the main collaborators, 
who the author cites most, who cites the author most, etc.
Similarly, these informations can be extracted and compared
for a group of authors. These detailed reports would be a good target for a future improvement of the \inspire author profiles.

\subsection{Author metrics: correlations}\label{sec:corr}
The right panel of \fig{Correlations} shows the correlations among the metrics for authors
within the whole \inspire database. The metrics are:
number of papers $N^{\rm pap}_A$, number of citations $N^{\rm cit}_A$, $h$-index squared $h_A^2$,
number of individual papers $N^{\rm ipap}_A$, number of individual citations $N^{\rm icit}_A$, PaperRank of authors $R_A$, AuthorRank of authors $\mathscr{R}_A$, and Citation-coin $\CC_A$. 
We consider the square of the $h$ index since this is known to be almost fully correlated ($0.99$) with the number of citations \citep{Hirsch:2005zc}.
From the table we see that our metrics for authors differ strongly
from the traditional ones, and also mildly differ between them. 
The main difference arises because of the difference between experimentalists and theorists,
so that the metrics become more correlated if restricted within each group. However, the combined effect of dividing by the number of references of the citing paper and the number of co-authors of the cited one makes our proposed bibliometric indicators fairly uncorrelated with the existing ones.

\begin{table}
\tiny
\begin{center}\renewcommand{\arraystretch}{0.9}\setlength\tabcolsep{3pt}
\begin{tabular}{clc}
 & Institution & All \\  \hline 
\color{red}\bf 1& \color{red}\bf CERN & \color{red}\bf 25065\\
\color{magenta}\bf 2& \color{magenta}\bf Princeton U. & \color{magenta}\bf 10774\\
\color{brown}\bf 3& \color{brown}\bf Harvard U. & \color{brown}\bf 10702\\
\bf 4& \bf SLAC & \bf 9913\\
\bf 5& \bf Princeton, Inst & \bf 9699\\
\bf 6& \bf Fermilab & \bf 9536\\
\bf 7& \bf Caltech & \bf 8470\\
\bf 8& \bf MIT, LNS & \bf 7658\\
\bf 9& \bf Brookhaven & \bf 7149\\
\bf 10& \bf LBL, Berkeley & \bf 6353\\
11& Cambridge U. & 5586\\
12& SUNY, Stony Bro & 5371\\
13& UC, Berkeley & 4838\\
14& DESY & 4580\\
15& Moscow, ITEP & 4495\\
16& Dubna, JINR & 4431\\
17& Washington U.,  & 4089\\
18& Munich, Max Pla & 4024\\
19& Maryland U. & 3801\\
20& Cornell U., LNS & 3786\\
21& Los Alamos & 3573\\
22& Oxford U. & 3570\\
23& Chicago U., EFI & 3569\\
24& Columbia U. & 3560\\
25& Imperial Coll., & 3414\\
26& Pennsylvania U. & 3304\\
27& UCLA & 3200\\
28& KEK, Tsukuba & 3164\\
29& Wisconsin U., M & 3101\\
30& Stanford U., Ph & 3080\\
31& Saclay & 3054\\
32& Texas U. & 2941\\
33& UC, Santa Barba & 2920\\
34& Santa Barbara,  & 2906\\
35& Yale U. & 2887\\
36& Heidelberg U. & 2847\\
37& Harvard-Smithso & 2633\\
38& Lebedev Inst. & 2609\\
39& Illinois U., Ur & 2586\\
40& Argonne & 2559\\
41& Kyoto U. & 2445\\
42& Rutgers U., Pis & 2441\\
43& Tokyo U. & 2430\\
44& Durham U. & 2397\\
45& St. Petersburg, & 2385\\
46& Cambridge U., D & 2343\\
47& Moscow, INR & 2234\\
48& ICTP, Trieste & 2183\\
49& Bohr Inst. & 2163\\
50& Minnesota U. & 2139\\
51& Utrecht U. & 2117\\
52& Frascati & 2097\\
53& Novosibirsk, IY & 2096\\
54& Landau Inst. & 2034\\
55& Michigan State  & 2033\\
56& Ohio State U. & 2021\\
57& Michigan U. & 1963\\
58& Rutherford & 1951\\
59& Weizmann Inst. & 1950\\
60& Rome U. & 1897\\
61& MIT & 1880\\
62& Munich, Tech. U & 1865\\
63& Serpukhov, IHEP & 1855\\
64& Penn State U. & 1813\\
65& Orsay, LPT & 1804\\
66& Ecole Normale S & 1791\\
67& Hamburg U. & 1777\\
68& Syracuse U. & 1756\\
69& Indiana U. & 1755\\
70& UC, San Diego & 1717\\
71& Rochester U. & 1714\\
72& Carnegie Mellon & 1687\\
73& Beijing, Inst.  & 1683\\
74& Garching, Max P & 1665\\
75& Chicago U. & 1656\\
76& Frankfurt U. & 1644\\
77& Tel Aviv U. & 1604\\
78& Texas A-M & 1596\\
79& Zurich, ETH & 1587\\
80& Bern U. & 1584
\end{tabular}~~
\begin{tabular}{lccccccccc}
  Institution & After 2010 & \!\!hep-ex\!\! & \!\!hep-ph \!\!& \!\!hep-th \!\!& \!\!gr-qc \!\!& \!\!astro-ph \!\! & \!\! hep-lat \!\! & \!\! nucl-th\!\! & \!\!nucl-ex\!\! \\  \hline 
\color{red}\bf CERN & \color{red}\bf 2817 & \color{red}\bf413 & \color{red}\bf652 & \bf184 & 10 & 49 & \bf66 & 8 & \color{red}\bf84\\
\color{magenta}\bf Fermilab & \color{magenta}\bf 1346 & \color{magenta}\bf227 & \color{brown}\bf289 & 3 & 3 & \bf111 & 23 & 2 & 10\\
\color{brown}\bf Brookhaven & \color{brown}\bf 791 & \bf77 & \bf202 & 7 & 0 & 15 & \color{brown}\bf103 & \bf54 & \color{magenta}\bf73\\
\bf Perimeter Inst. & \bf 751 & 0 & 81 & \color{magenta}\bf434 & \color{red}\bf133 & 35 & 2 & 0 & 0\\
\bf DESY & \bf 705 & \bf147 & \bf274 & 70 & 1 & 38 & 10 & 1 & 9\\
\bf Princeton, Inst & \bf 686 & 0 & 41 & \color{red}\bf522 & 6 & 90 & 0 & 0 & 0\\
\bf Dubna, JINR & \bf 639 & \bf124 & 137 & 55 & 9 & 12 & 6 & 34 & \color{brown}\bf53\\
\bf KEK, Tsukuba & \bf 629 & \bf99 & 70 & 37 & 6 & 24 & 45 & 5 & 5\\
\bf SLAC & \bf 603 & 71 & 177 & 47 & 2 & 72 & 0 & 1 & 5\\
\bf Caltech & \bf 584 & 40 & 53 & 159 & \bf93 & \color{brown}\bf166 & 1 & 2 & 4\\
Beijing, Inst.  & 581 & \color{brown}\bf204 & 125 & 18 & 5 & 69 & 4 & 9 & 4\\
Princeton U. & 554 & 39 & 37 & \bf232 & 24 & \bf132 & 0 & 0 & 5\\
Munich, Max Pla & 525 & 56 & \bf191 & 123 & 2 & 53 & 12 & 0 & 6\\
LBL, Berkeley & 519 & 57 & 84 & 27 & 1 & 80 & 4 & 28 & \bf37\\
Valencia U., IF & 492 & 52 & \color{magenta}\bf304 & 7 & 14 & 36 & 7 & 17 & 1\\
Imperial Coll., & 488 & \bf80 & 37 & \bf179 & 25 & 42 & 1 & 1 & 6\\
Cambridge U., D & 460 & 0 & 29 & \color{brown}\bf262 & \bf79 & 63 & 16 & 3 & 0\\
Tokyo U., IPMU & 416 & 27 & 116 & 165 & 15 & 67 & 2 & 1 & 0\\
Heidelberg, Max & 400 & 33 & \bf221 & 1 & 0 & 68 & 0 & 3 & 9\\
Stanford U., Ph & 383 & 14 & 20 & \bf215 & 16 & 55 & 0 & 0 & 4\\
INFN, Rome & 365 & \bf84 & 68 & 12 & 32 & 61 & 10 & 5 & 17\\
APC, Paris & 359 & 22 & 9 & 57 & \bf72 & \color{red}\bf170 & 0 & 0 & 0\\
UC, Berkeley & 358 & 15 & 79 & 70 & 7 & 103 & 4 & 3 & 5\\
Maryland U. & 352 & 24 & 87 & 35 & 49 & 74 & 14 & 6 & 6\\
Heidelberg U. & 352 & 63 & 85 & 44 & 4 & 37 & 10 & 18 & 23\\
Potsdam, Max Pl & 348 & 0 & 3 & 166 & \color{magenta}\bf132 & 30 & 0 & 1 & 0\\
MIT & 348 & 65 & 35 & 37 & 18 & 35 & 4 & 26 & 24\\
Munich, Tech. U & 344 & 24 & \bf196 & 9 & 0 & 23 & 4 & 26 & 12\\
Washington U.,  & 338 & 21 & 57 & 48 & 6 & 22 & \bf67 & 41 & 12\\
Harvard U., Phy & 334 & 15 & 48 & \bf189 & 2 & 16 & 0 & 0 & 0\\
Jefferson Lab & 333 & 10 & 119 & 0 & 0 & 0 & \bf48 & 22 & \bf41\\
Wisconsin U., M & 332 & 61 & 102 & 23 & 0 & 96 & 0 & 3 & 6\\
Zurich, ETH & 332 & 45 & 66 & 79 & 6 & 31 & 13 & 0 & 5\\
Zurich U. & 324 & 40 & \bf190 & 2 & 12 & 56 & 0 & 0 & 4\\
Frascati & 324 & \bf81 & 46 & 13 & 3 & 3 & 6 & 0 & 22\\
Oxford U. & 315 & 75 & 35 & 14 & 12 & \bf108 & 0 & 3 & 2\\
Durham U., IPPP & 314 & 2 & \bf267 & 14 & 0 & 18 & 0 & 0 & 0\\
NIKHEF, Amsterd & 312 & 61 & 114 & 18 & 21 & 23 & 0 & 0 & 18\\
SUNY, Stony Bro & 310 & 35 & 56 & 66 & 0 & 29 & 7 & \bf57 & 23\\
UC, Santa Barba & 306 & 23 & 17 & \bf180 & 21 & 28 & 6 & 0 & 3\\
Moscow, ITEP & 305 & 55 & 72 & 66 & 2 & 6 & 13 & 7 & \bf33\\
Frankfurt U. & 304 & 4 & 85 & 7 & 20 & 17 & \bf49 & \color{magenta}\bf91 & 12\\
Kyoto U., Yukaw & 302 & 0 & 36 & 132 & 32 & 44 & 16 & 25 & 0\\
UCLA & 301 & 19 & 33 & 105 & 2 & 77 & 1 & 3 & 15\\
INFN, Pisa & 296 & 63 & 45 & 25 & 16 & 44 & 15 & 13 & 3\\
Tokyo U. & 294 & 19 & 78 & 43 & 9 & 25 & 23 & 22 & 15\\
Ohio State U. & 286 & 28 & 61 & 24 & 0 & 40 & 12 & \color{brown}\bf89 & 10\\
Columbia U. & 285 & 19 & 37 & 48 & 5 & 91 & 31 & 10 & 9\\
McGill U. & 284 & 14 & 87 & 80 & 8 & 46 & 1 & \bf43 & 0\\
Kyoto U. & 282 & 20 & 36 & 77 & 23 & 51 & 12 & 23 & 6\\
INFN, Turin & 281 & 41 & 88 & 27 & 4 & 27 & 6 & 7 & \bf36\\
Bonn U. & 279 & 55 & 93 & 27 & 0 & 16 & 11 & 10 & 8\\
Cracow, INP & 278 & 47 & 134 & 0 & 0 & 6 & 0 & \bf57 & 12\\
Harvard U. & 278 & 5 & 35 & 177 & 6 & 33 & 0 & 0 & 0\\
Madrid, IFT & 274 & 1 & 105 & 96 & 9 & 38 & 8 & 3 & 0\\
Los Alamos & 271 & 20 & 78 & 3 & 3 & 26 & 18 & 33 & 22\\
UC, Irvine & 267 & 38 & 100 & 6 & 0 & 83 & 0 & 0 & 0\\
Orsay, LPT & 265 & 1 & 179 & 30 & 26 & 3 & 13 & 0 & 0\\
Tata Inst. & 264 & 22 & 36 & 66 & 40 & 10 & 20 & 20 & 6\\
Weizmann Inst. & 258 & 16 & 55 & 80 & 3 & 29 & 1 & 1 & 6\\
Southampton U. & 253 & 1 & 134 & 48 & 26 & 13 & 25 & 0 & 0\\
Stanford U., IT & 253 & 1 & 45 & \bf180 & 1 & 14 & 0 & 0 & 0\\
Pennsylvania U. & 252 & 22 & 5 & 111 & 3 & 64 & 0 & 0 & 2\\
Bohr Inst. & 252 & 18 & 52 & 103 & 6 & 33 & 14 & 3 & 12\\
Aachen, Tech. H & 247 & 66 & 73 & 1 & 1 & 49 & 0 & 0 & 8\\
Yale U. & 246 & 29 & 16 & 32 & 1 & 39 & 5 & 26 & 25\\
Moscow, INR & 241 & 39 & 43 & 23 & 4 & 47 & 0 & 3 & 18\\
Darmstadt, GSI & 240 & 7 & 36 & 1 & 0 & 13 & 5 & 42 & \bf26\\
Saclay, SPhT & 237 & 0 & 102 & 75 & 3 & 19 & 0 & 27 & 2\\
INFN, Padua & 237 & 49 & 58 & 24 & 6 & 37 & 0 & 1 & 20\\
Argonne & 236 & 33 & 83 & 2 & 0 & 12 & 5 & 15 & 11\\
Manchester U. & 235 & 65 & 73 & 12 & 0 & 50 & 0 & 6 & 2\\
Johns Hopkins U & 233 & 16 & 64 & 19 & 3 & \bf122 & 0 & 0 & 2\\
INFN, Trieste & 230 & 21 & 45 & 45 & 23 & 43 & 0 & 0 & 7\\
British Columbi & 229 & 18 & 15 & 79 & 17 & 68 & 0 & 1 & 2\\
Michigan State  & 226 & 27 & 106 & 1 & 0 & 17 & 6 & 24 & 12\\
Texas A-M & 225 & 31 & 43 & 50 & 2 & 25 & 0 & 29 & 18\\
Humboldt U., Be & 225 & 15 & 67 & 67 & 2 & 13 & 46 & 0 & 0\\
Rome U. & 224 & 22 & 47 & 13 & 37 & 61 & 6 & 4 & 4\\
MIT, Cambridge, & 218 & 0 & 93 & 85 & 0 & 14 & 16 & 4 & 0
\end{tabular}
\end{center}
\caption{\em\label{tab:Instit} Institutions listed according to their contribution to fundamental physics
(as defined by {\sc InSpire})
quantified as the number of individual citations received by their affiliates 
as defined in \eq{NictiI}.
Left: all time.  Right: from 2010, and split within arXiv categories.
The top ten in each category are highlighted.}
\end{table}

\section{Rankings groups}\label{sec:GeoRank}
Our metrics respect sum rules and thereby allow to define metrics for groups by simply summing over their members.
Furthermore, the main property of the Citation-coin holds not only for authors of a set of papers,
but for any group: it cannot be increased trough internal citations. In order to show illustrative results we mostly
use the number of individual citations as the metric to rank groups, since individual citations have the benefit of being very fast to compute. Of course, one could equally apply any other metric discussed before. 

\subsection{Ranking institutions}\label{sec:inst}
We share the number of citations received by each paper between the institutions $I$ of its authors 
with weights $p_I$ that sum up to 1.
The weights are computed by first sharing equally each paper
between its  $N_{\rm aut}$ authors,
and next between the affiliations of each author.
Thereby each author $A$ contributes to institute $I$ as $1/(N_A^{\rm aff} N_p^{\rm aut})$, 
where $N_A^{\rm aff}$ is the number of affiliations of author $A$, that include institute $I$.
When some affiliation is missed by {\rm InSpire}, we renormalize the $p_I$ such 
that they sum up to 1.
Next, we sum over all papers $p$ (optionally restricting to recent papers, if one wants to evaluate the present situation, rather than the all-time record).
 In formul\ae, the number of individual citations received by institute $I$ is:
\beq \label{eq:NictiI}
N^{\rm icit}_{I} = \sum_{p} p_{pI} N_p^{\rm icit}\,,\qquad p_{pI} = \frac{1}{N_p^{\rm aut} }\sum_{A\in I,p} \frac{1}{N^{\rm aff}_{A}}\,.\eeq
In the left column of \tabl{Instit} we list the institutes that most contributed to fundamental physics, according to the whole
\inspire database.
In order to focus on the present situation,
in the second column we restrict to recent papers, written from 2010.
The top positions are occupied by research institutions rather than by teaching institutions.
In the right columns of \tabl{Instit} we show the contributions
within each main arXiv category:
the best institutions strongly depend on the sub-field of interest.
This means that generic rankings (e.g.\ at the faculty level)
are not much useful for authors interested in finding the most active institutions within their specific sub-fields.

\smallskip

\begin{figure}[t]
$$\includegraphics[width=0.56\textwidth]{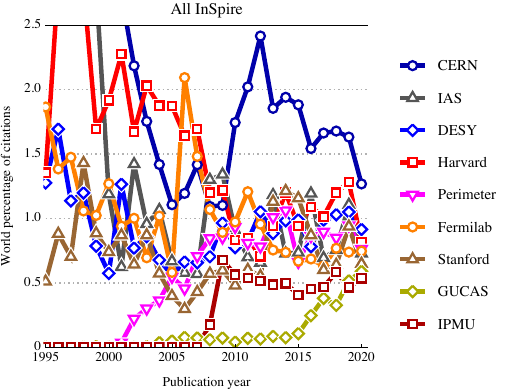}\quad
\includegraphics[width=0.42\textwidth]{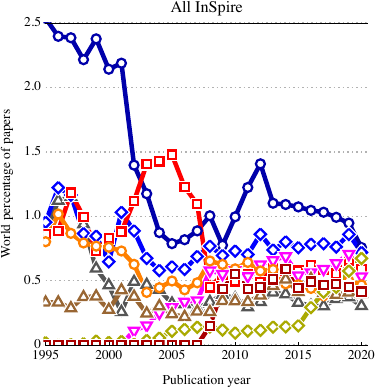}$$
\caption{\label{fig:InstEvo}\em Percentage impact of some main institutes
according to citations (left) and papers (right).}
\end{figure}

Concerning the time evolution, we compute the percentage of individual citations received by papers written within 
any given year by authors affiliated to each institute, and fully fractionalized according to \eq{NictiI}.
\fig{InstEvo}a shows how the percentage impact of some main institutes evolved with time.
As papers published recently will accumulate most of their citations in the future, in 
\fig{InstEvo}b we repeat the analysis in terms of fractionally counted papers.
This indicator provides a less significant proxy for scientific merit than citations,
but its value is immediately available, and helps to interpret the evolution in citations.

\begin{figure}[t]
$$\includegraphics[height=0.3\textwidth]{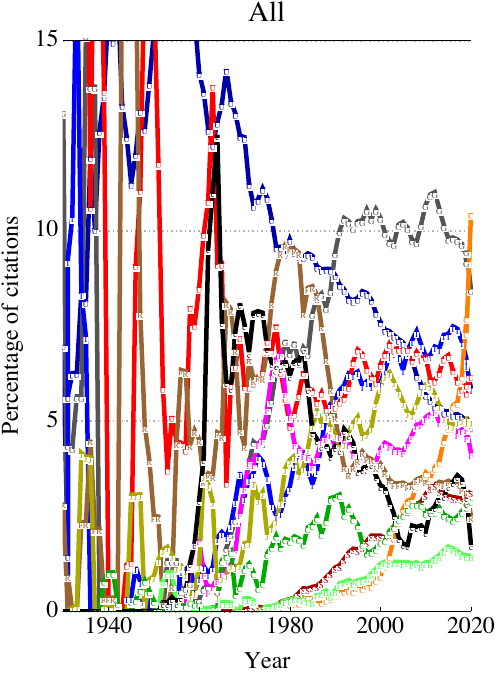}\quad
\includegraphics[height=0.3\textwidth]{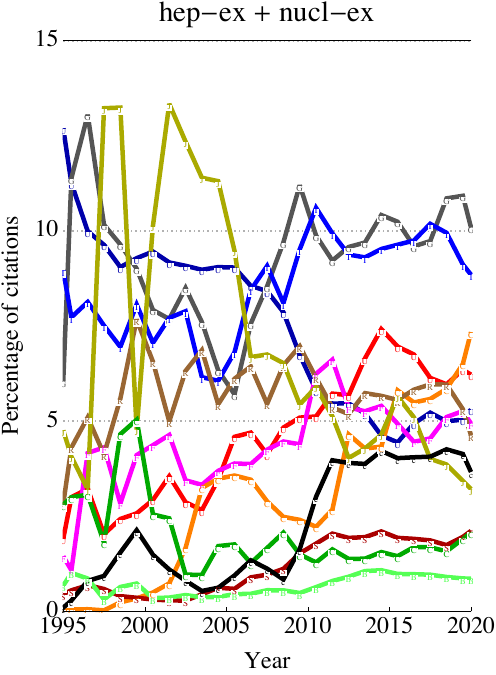}\quad
\includegraphics[height=0.3\textwidth]{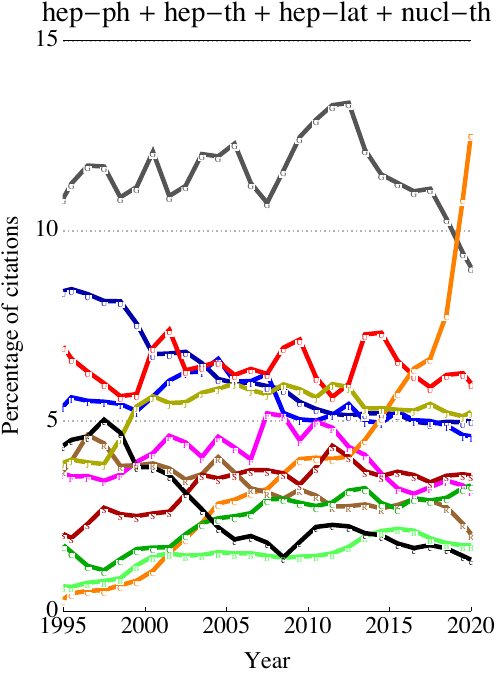}\quad
\includegraphics[height=0.3\textwidth]{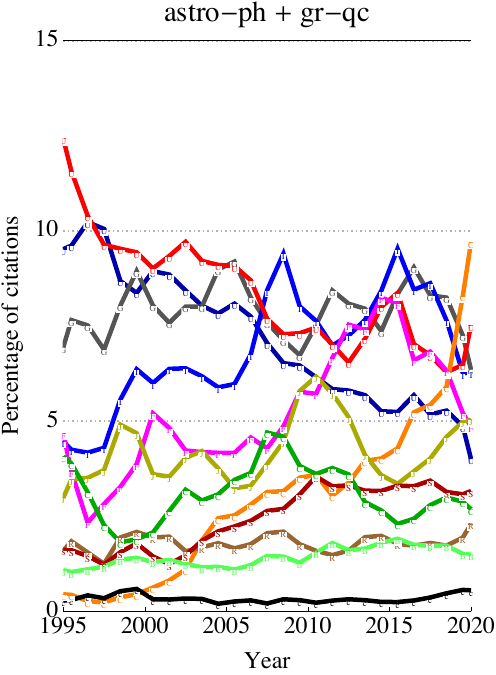}$$
$$\includegraphics[width=0.99\textwidth]{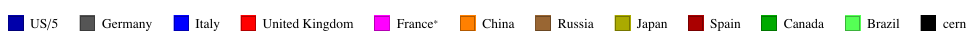}$$
\caption{\label{fig:countries}\em Percentage impact of representative countries.
For ease of visualisation, the USA contribution has been multiplied by 1/5.
 The $*$ on France means that CERN is plotted separately.}
 \via{
$$\includegraphics[height=0.3\textwidth]{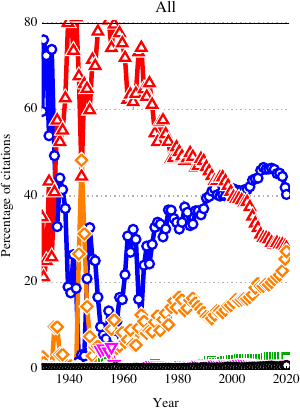}\
\includegraphics[height=0.3\textwidth]{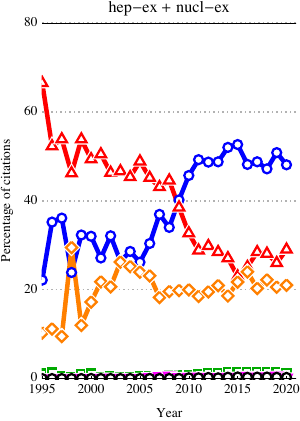}\quad
\includegraphics[height=0.3\textwidth]{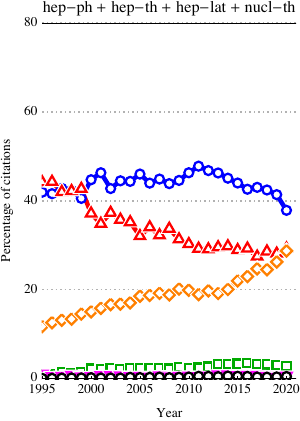}\quad
\includegraphics[height=0.3\textwidth]{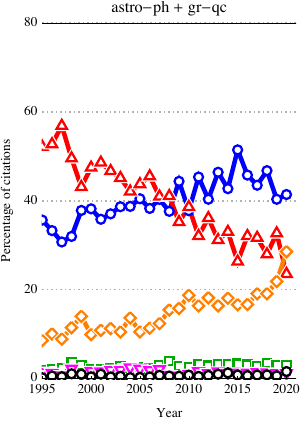}$$
$$\includegraphics[width=0.7\textwidth]{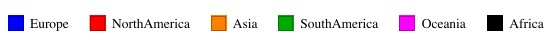}$$
\caption{\label{fig:continent}\em Percentage impact by continent.}}
\end{figure}

The black curve in \fig{countries} shows the time-dependence of the contribution of the top institution, CERN.
It reached a maximum around 1965 (12\% of world-wide individual citations to 1965 papers  have been given to CERN authors) 
and declined stabilising to $\approx2\%$.
All main historical  institutions show a similar trend, due to the fact that,
until 1970, fundamental physics was concentrated in a few institutions,
and later became more distributed (especially in theoretical physics).
Half of the impact, quantified using individual citations, was produced in the 12 top institutions in 1970,
in 22 in 1980, 42 in 1990, 80 in 2000 and 160 now.
As a consequence the relative impact of the top institutions declined.
Among the new institutions, Perimeter and IPMU reached very high positions.

\begin{table}[t]
\begin{center}\footnotesize\renewcommand{\arraystretch}{0.9}\setlength\tabcolsep{9pt}
\begin{tabular}{clc|ccc} \\ & Institution & $N_{\rm iaut}$ & $N_{\rm icit}$ & PaperRank $R$ & AuthorRank $\mathscr{R}$ \\  \hline 
1& Princeton, Inst. Adv & 32 & 2.1 \% & 2.2 \% & 2.4 \% \\
2& CERN & 896 & 1.7 \% & 2.0 \% & 2.7 \% \\
3& Fermilab & 435 & 1.4 \% & 1.3 \% & 1.8 \% \\
4& Texas U. & 46 & 1.0 \% & 3.6 \% & 1.6 \% \\
5& Caltech & 90 & 0.89 \% & 3.6 \% & 1.6 \% \\
6& Brookhaven & 218 & 0.83 \% & 0.97 \% & 0.94 \% \\
7& Cambridge U., DAMTP & 49 & 0.78 \% & 0.96 \% & 0.79 \% \\
8& SLAC & 141 & 0.73 \% & 0.83 \% & 0.90 \% \\
9& Dubna, JINR & 312 & 0.62 \% & 0.54 \% & 0.51 \% \\
10& DESY & 276 & 0.59 \% & 0.46 \% & 0.57 \% \\
11& Princeton U. & 74 & 0.59 \% & 0.74 \% & 0.71 \% \\
12& Harvard U., Phys. De & 35 & 0.58 \% & 0.89 \% & 0.74 \% \\
13& Maryland U. & 72 & 0.55 \% & 0.45 \% & 0.46 \% \\
14& Imperial Coll., Lond & 121 & 0.54 \% & 0.56 \% & 0.55 \% \\
15& Perimeter Inst. Theo & 70 & 0.53 \% & 0.38 \% & 0.41 \% \\
16& IPhT, Saclay & 40 & 0.52 \% & 0.53 \% & 0.50 \% \\
17& Stanford U., ITP & 22 & 0.51 \% & 0.69 \% & 0.62 \% \\
18& LBL, Berkeley & 116 & 0.49 \% & 0.68 \% & 0.60 \% \\
19& Munich, Max Planck I & 124 & 0.47 \% & 0.36 \% & 0.44 \% \\
20& Chicago U., EFI & 41 & 0.46 \% & 0.60 \% & 0.54 \% 
\end{tabular}
\end{center}
\caption{\em\label{tab:InstitByAuthor} Institutes listed according to all-time bibliometric ranks
of their last-year affiliates.
For each institute, we show the number of individual authors
active in the last year, and the sum of their biblio-metric indicators (percentage of the world total):
number of individual citations, PaperRank of authors, and AuthorRank of authors.}
\end{table}

The list in \tabl{Instit} highlights the institutes with the most productive authors in recent times
(often young authors).
The list in \tabl{InstitByAuthor}  highlights institutes with the most productive affiliates,
as instead quantified by their all-time biblio-metric rankings.
\Tabl{InstitByAuthor} is produced as follows.
Since no list of present affiliates is available,  
we use the declared affiliations of authors that wrote at least one paper in the last year, 2020.
Authors with $N_{\rm aff}$ affiliations are assigned with fraction
$1/N_{\rm aff}$ to each affiliation.
When this number differs in recent papers, we average over them respecting sum rules:
each authors is affiliated to various institutions with percentages that sum up to one. 
The average suppresses minor mistakes/missing data in {\sc InSpire}.
For each institute $I$, we obtain a list of active affiliates with their percentages.
Summing over these percentages we determine the number of `individual authors' $N_{\rm iaut}$
affiliated to each institution,
shown in the 3rd column of \tabl{InstitByAuthor}.
Next, summing over all affiliates using the same weights, we compute the total biblio-metric ranking of all authors in each institute.
In column 4 we show the all-time number of individual citations,
in column 5 the PaperRank (\secc{rank})
and in column the AuthorRank (\secc{authorrank}).
The latter 3 columns actually show the world percentage of each metric in the various institutes:
about $2\%$ of researchers that most contributed to fundamental physics can be found at IAS in Princeton, or at CERN.

\begin{figure}[t]
$$\includegraphics[width=0.98\textwidth]{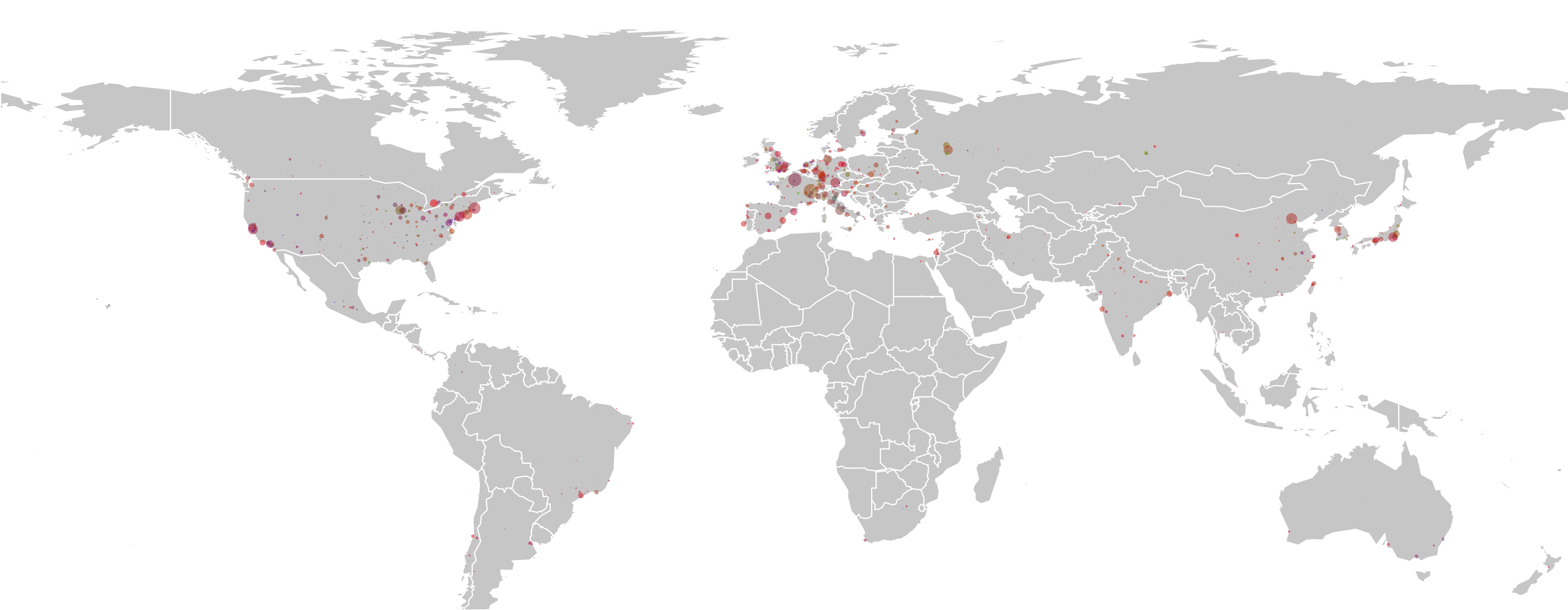}$$
\caption{\label{fig:MapInst}\em Institutes that contributed to fundamental physics plotted as dots with area proportional to the number of individual citations received by their papers written from year 2010.
We group institutes closer than $30\,{\rm km}$.
The amount of green,  red, blue in the color is proportional to the contribution to experiment, theory, astro-cosmology respectively.}
\end{figure}

\begin{figure}[t]
$$\includegraphics[width=\textwidth]{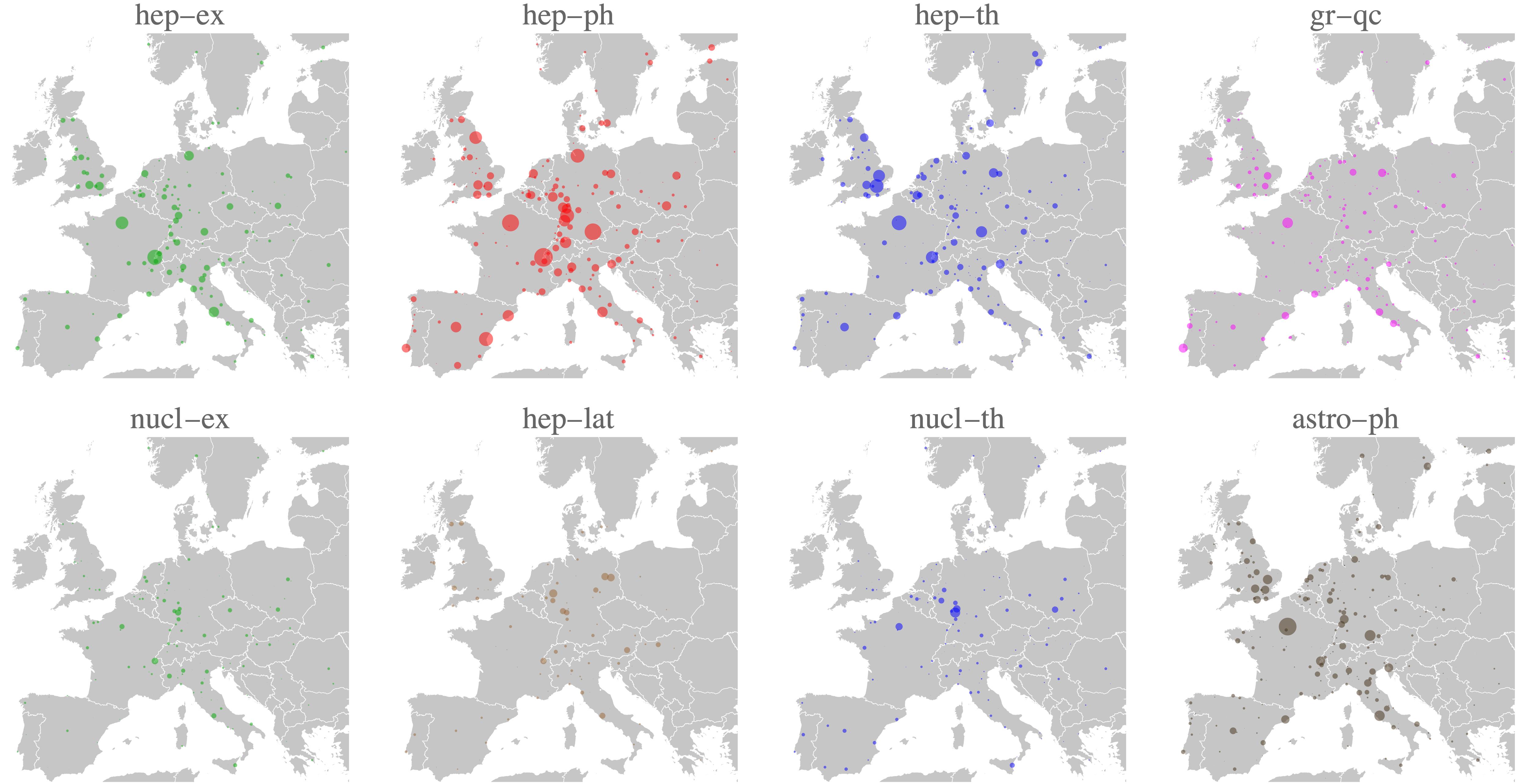}$$
\caption{\label{fig:MapInstEu}\em As in \fig{MapInst}, restricted to Europe, and showing separately
the contributions within main arXiv representative categories.}
\end{figure}

\subsection{Ranking towns}\label{sec:towns}
Sometimes multiple institutes are located nearby, and what matters is their total.
We group together institutes closer than about $30\,{\rm km}$.
In \fig{MapInst} we show a map with the places that mostly contributed to fundamental physics:
each contribution is plotted as a circle with area 
proportional to the number of individual citations received by their papers written from year 2010,
and color proportional to the contribution to experiment (green), theory (red), astro-cosmology (blue) respectively.
We focus on a relatively recent period, such that the map photographs the present situation.

Similar maps can be computed for any given sub-topic or region.
For instance, \fig{MapInstEu} shows the same map separated according to 
papers published within the main arXiv categories, and restricted to Europe.

\via{
\begin{figure}[t]
$$\includegraphics[width=0.9\textwidth]{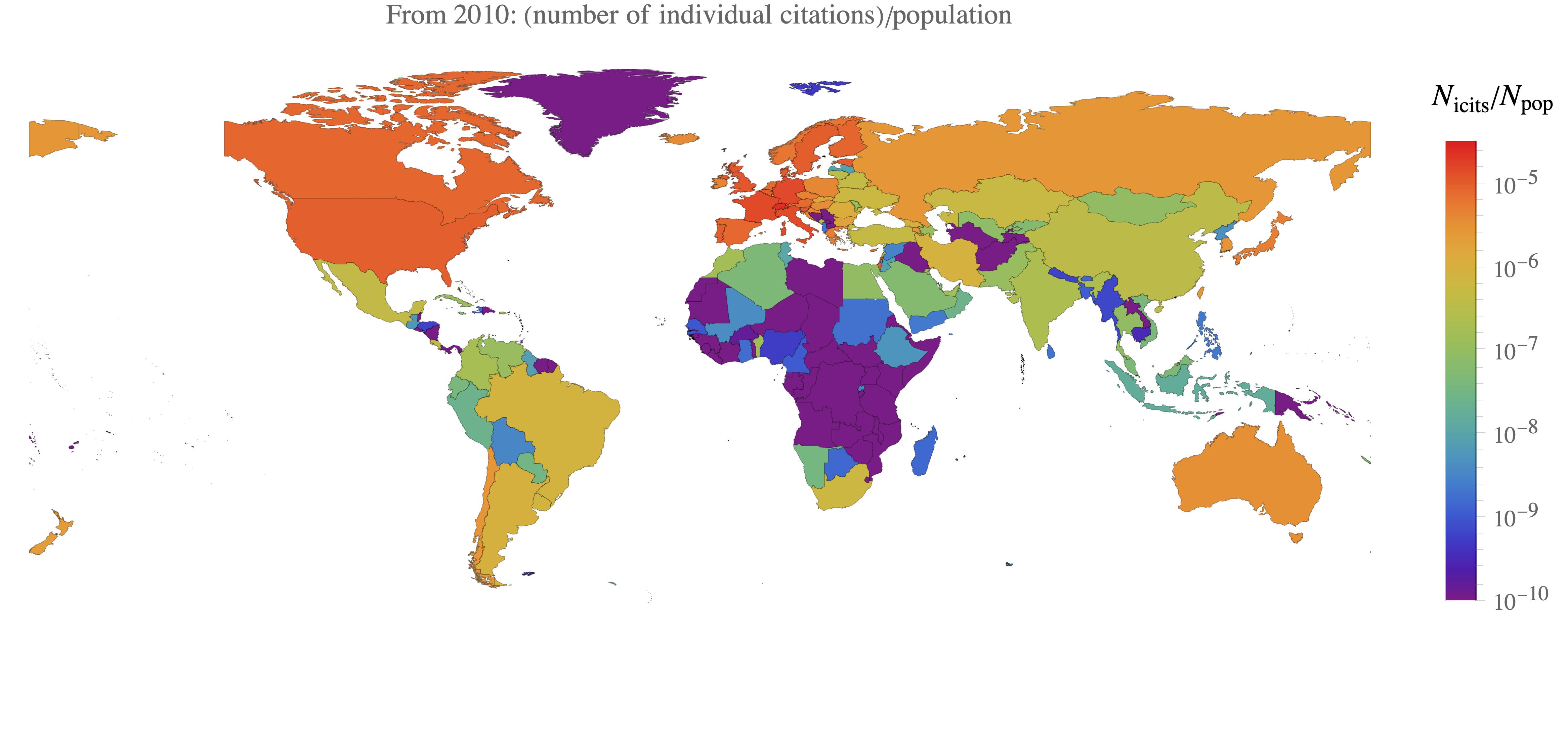}$$
\vspace{-2cm}
\caption{\label{fig:MapCitPop}\em Individual citations per country from 2010 divided by population.
The top countries are Switzerland, Germany, France, Italy, Slovenia.}
$$\includegraphics[width=0.9\textwidth]{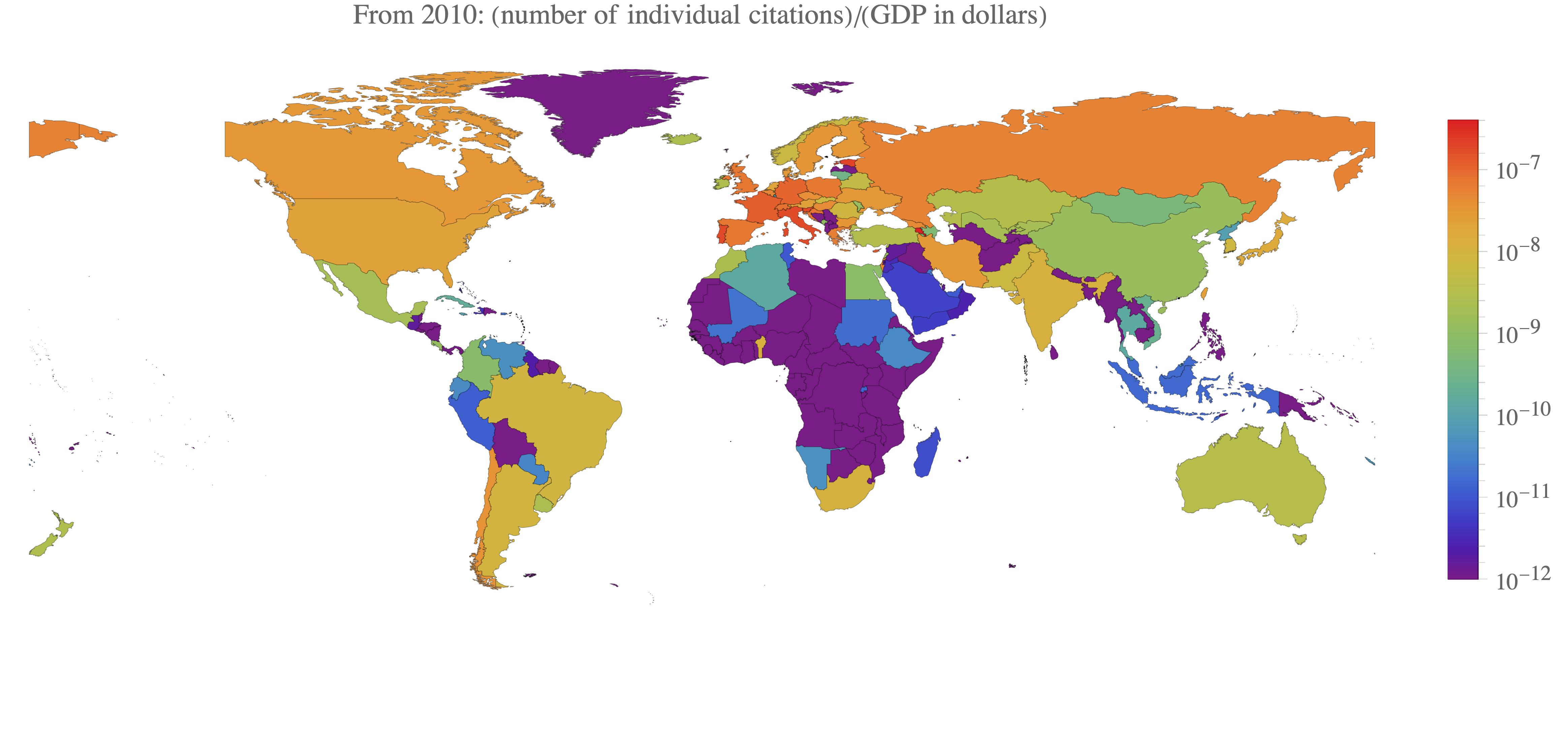}$$
\vspace{-2cm}
\caption{\label{fig:MapCitGDP}\em Individual citations per country from 2010 divided by domestic gross product in dollars.
The top countries are Armenia, Estonia, Slovenia, Portugal, Italy.}
\end{figure}}

\subsection{Ranking countries\via{ and continents}}\label{sec:country}
We rank a country \via{or continent} $C$ by summing the ranks of all institutes located in $C$.
We apply this to the number of individual citations:
\beq N^{\rm icit}_{C} = \sum_{I\in C} N^{\rm icit}_{I}\,.\eeq
In the left panel of \fig{countries} we show the  time evolution of the impact of papers written in
main representative countries within each year.
The impact is quantified as the percentage of the world total,
in order to factor out the reduced number of citations of recent papers.
USA is the main country, but declining (from $70\%$ around 1950 to 25\% now); 
European countries are now stable or slightly growing;
China is growing.
In the right panels of \fig{countries} the time evolution of the percentage contribution of each country
is shown separately, after the advent of arXiv, within the main fields: experiment, theory, astro-cosmology.

\smallskip
\via{
\Fig{continent} shows the analogous plot for continents.
We see that European physics suffered a big decline after WW2, and returned to be the main actor only around 2000.
The decline of Asia around 1985 is due to the fall of Soviet Union (Mathematica geographic tools assign all Russia to Asia);
the present rise of Asia is mostly due to China.}

\medskip

\via{\Fig{MapCitPop} shows the ratio between the number of  individual citations and the population of countries, while \fig{MapCitGDP} shows the ratio between the number of individual citations and the gross domestic product. }

\begin{table}[t]  
\begin{center}\footnotesize\renewcommand{\arraystretch}{0.9}\setlength\tabcolsep{2pt}
\begin{tabular}{clccc}
 & Journal, all \inspire& $N_{\rm icit}$  & $N_{\rm icit}/N_{\rm pap}$  & $\CC$  \\  \hline 
1& Phys.Rev.D & 113601 & \color{verdes}1.6 & $\color{verdes}41080$ \\
2& Phys.Lett.B & 80214 & \color{verdes}1.5 & $\color{verdes}25748$ \\
3& Phys.Rev.lett. & 67864 & \color{verdes}2.9 & $\color{verdes}44359$ \\
4& Nucl.Phys.B & 56150 & \color{verdes}2.3 & $\color{verdes}31339$ \\
5& Astrophys.J. & 38960 & \color{verdes}1.3 & $\color{verdes}9920$ \\
6& Phys.Rev.C & 30386 & \color{red}0.9 & $\color{red}-4071$ \\
7& JHEP & 30044 & \color{verdes}1.5 & $\color{verdes}10024$ \\
8& Nucl.Phys.A & 21518 & \color{red}0.7 & $\color{red}-9310$ \\
9& Mon.Not.Roy.Astron.S & 20594 & \color{red}0.9 & $\color{red}-2525$ \\
10& Nucl.Instrum.Meth.A & 17592 & \color{verdes}1.0 & $\color{verdes}380$ \\
11& Phys.Rev. & 17351 & \color{verdes}2.4 & $\color{verdes}10185$ \\
12& Astron.Astrophys. & 14323 & \color{red}0.8 & $\color{red}-4762$ \\
13& Astrophys.J.Lett. & 10703 & \color{verdes}1.1 & $\color{verdes}1351$ \\
14& Eur.Phys.J.C & 9964 & \color{verdes}1.4 & $\color{verdes}2801$ \\
15& Phys.Rept. & 9788 & \color{verdes}7.4 & $\color{verdes}8462$ \\
16& Class.Quant.Grav. & 8822 & \color{red}0.9 & $\color{red}-731$ \\
17& Commun.Math.Phys. & 7792 & \color{verdes}1.9 & $\color{verdes}3658$ \\
18& Annals Phys. & 7765 & \color{verdes}2.2 & $\color{verdes}4169$ \\
19& Rev.Mod.Phys. & 7235 & \color{verdes}4.8 & $\color{verdes}5732$ \\
20& Z.Phys.C & 6939 & \color{verdes}1.4 & $\color{verdes}1820$ \\
21& J.Math.Phys. & 5717 & \color{red}0.7 & $\color{red}-2289$ \\
22& Astron.J. & 5581 & \color{verdes}1.1 & $\color{verdes}710$ \\
23& JCAP & 5331 & \color{verdes}1.1 & $\color{verdes}509$ \\
24& Astrophys.J.Suppl. & 4824 & \color{verdes}2.9 & $\color{verdes}3148$ \\
25& Nature & 4406 & \color{verdes}3.0 & $\color{verdes}2945$ \\
26& Prog.Theor.Phys. & 4385 & \color{red}0.6 & $\color{red}-2697$ \\
27& Int.J.Mod.Phys.A & 4374 & \color{red}0.5 & $\color{red}-4459$ \\
28& Yad.Fiz. & 4072 & \color{red}0.4 & $\color{red}-5218$ \\
29& Comput.Phys.Commun. & 3864 & \color{verdes}2.6 & $\color{verdes}2373$ \\
30& JINST & 3816 & \color{verdes}1.5 & $\color{verdes}1330$ 
\end{tabular}~~\begin{tabular}{lccc}
  Journal, after 2010& $N_{\rm icit}$  & $N_{\rm icit}/N_{\rm pap}$  & $\CC$  \\  \hline 
Phys.Rev.D & 19288 & \color{verdes}1.0 & $\color{verdes}656$ \\
JHEP & 13942 & \color{verdes}1.3 & $\color{verdes}3179$ \\
Phys.Rev.lett. & 9488 & \color{verdes}2.4 & $\color{verdes}5517$ \\
Phys.Lett.B & 6008 & \color{verdes}1.3 & $\color{verdes}1261$ \\
Astrophys.J. & 5929 & \color{red}0.6 & $\color{red}-3700$ \\
Phys.Rev.C & 5666 & \color{red}0.9 & $\color{red}-371$ \\
Mon.Not.Roy.Astron.S & 5386 & \color{red}0.5 & $\color{red}-5091$ \\
Eur.Phys.J.C & 4376 & \color{verdes}1.4 & $\color{verdes}1216$ \\
JCAP & 3174 & \color{red}1. & $\color{red}-105$ \\
Astron.Astrophys. & 2647 & \color{red}0.5 & $\color{red}-3068$ \\
Nucl.Instrum.Meth.A & 2291 & \color{red}0.7 & $\color{red}-898$ \\
Class.Quant.Grav. & 1896 & \color{red}0.8 & $\color{red}-497$ \\
JINST & 1890 & \color{red}0.9 & $\color{red}-268$ \\
Astrophys.J.Lett. & 1661 & \color{red}0.7 & $\color{red}-683$ \\
Pos & 1327 & \color{red}0.1 & $\color{red}-9215$ \\
Nucl.Phys.B & 1321 & \color{red}0.8 & $\color{red}-327$ \\
J.Phys.Conf.Ser. & 1183 & \color{red}0.2 & $\color{red}-6450$ \\
J.Phys.G & 1046 & \color{red}1. & $\color{red}-44$ \\
Phys.Rev.B & 961 & \color{red}0.5 & $\color{red}-967$ \\
Astrophys.J.Suppl. & 903 & \color{verdes}1.7 & $\color{verdes}386$ \\
Nucl.Phys.A & 847 & \color{red}0.5 & $\color{red}-1028$ \\
Nature & 737 & \color{verdes}2.9 & $\color{verdes}484$ \\
Eur.Phys.J.A & 729 & \color{red}0.7 & $\color{red}-252$ \\
Astropart.Phys. & 717 & \color{verdes}1.4 & $\color{verdes}196$ \\
Chin.Phys.C & 706 & \color{red}0.7 & $\color{red}-316$ \\
Phys.Rev.ST Accel.Be & 672 & \color{red}0.9 & $\color{red}-73$ \\
Comput.Phys.Commun. & 630 & \color{verdes}2.0 & $\color{verdes}314$ \\
Int.J.Mod.Phys.A & 622 & \color{red}0.4 & $\color{red}-1152$ \\
Phys.Rept. & 584 & \color{verdes}6.2 & $\color{verdes}490$ \\
Science & 548 & \color{verdes}3.4 & $\color{verdes}387$ 
\end{tabular}
\caption{\em\label{tab:topj}  Number of individual citations ($N_{\rm icit} = \sum N_{\rm cit}/N_{\rm ref}$), average number of individual citations per paper ($N_{\rm icit}/N_{\rm pap}$) and Citation-coin ($\CC=N_{\rm icit}-N_{\rm pap}$) for some top journals.
The analysis is restricted to fundamental physics as included in the \inspire database.
Left: all time. Right: only papers published from year 2010.
}
\end{center}
\end{table}

\subsection{Ranking journals}\label{sec:journals}
\begin{figure}[t]
$$\includegraphics[width=0.75\textwidth]{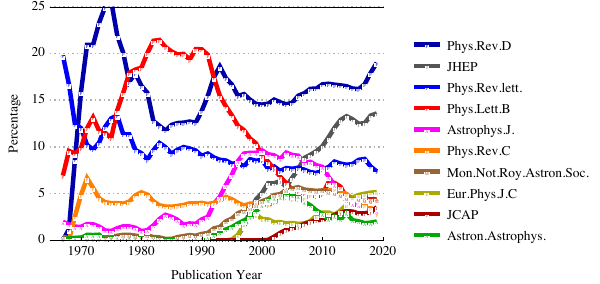}
$$
\caption{
\label{fig:JournalEvo}
\em 
Time evolution of the fraction of individual citations received by papers
published on some notable journals.
}
\end{figure}

When a paper is published on some journal (refereed or not), \inspire provides this information.
\Tabl{topj} lists journals according to the number of individual citations received by all papers they published in fundamental physics, as included in {\sc InSpire}.
We separately show these data for all {\sc InSpire}, and restricting to articles published from 2000.
\Fig{JournalEvo} shows the time evolution of the percentage number of individual citations
received by all papers on selected journals.
We see that  internet brought a revolution around 2000:
the decline of NPB and PLB and the emergence of JHEP and Astrophysics J. 
as preferred journals.

\smallskip

The 3rd column of \tabl{topj} shows a direct measure of `quality':
the average number of individual citations per paper, 
which roughly corresponds to what is known as `impact factor'.
According to this measure, the top-journals  are those that publish reviews.

\smallskip

The 4th column shows a measure of both `quality' and `quantity':
the Citation-coin $\CC$ of the journal 
(difference between the number of individual citations and the number of published papers).
The top journal according to this measure is PRL.
Journals that publish reviews score well, but are limited by their restricted scope.
The Citation-coin is negative for journals that publish many papers that do not attract much interest, 
in particular those that publish conference proceedings. 

\section{Conclusions}\label{sec:concl}
We applied improved bibliometric indicators to the \inspire database, which covers fundamental physics mostly after 1970.
\Fig{trends} shows some main trends: growing rate of
papers, growing number of authors and of references per paper.
\Fig{birthdeathrates} shows the health status of the main sub-fields.

The metrics that we explored are:
\begin{itemize}
\item The {\em number of individual citations} $N_{\rm icit}$ ---
defined as the number of citations divided by the numbers of authors of the cited paper and of references of citing papers ---
that compensates for the recent inflationary trends towards more authors and references.

\item The {\em PaperRank} $R$, which applies the PageRank algorithm to the citation network among papers
and that approximates a physical observable: how many times a paper is read.

\item The {\em AuthorRank} $\mathscr{R}$ which applies the PageRank algorithm to the individual citation network among authors.

\item The {\em Citation-coin} $\CC$ equal to the difference between the number of received and given individual citations.
Since a naive definition of this indicator, that cannot be increased by self-citations and circular citations, strongly penalizes young authors, we propose an improvement that takes into account the average time in which citations are received within the community.

\end{itemize}
An important feature of all these metrics is that they can be computed in practice.

In \secc{PaperRank} we apply these metrics to papers.
\Tabl{topcited0} shows the traditional list of all-time {\em top-cited} (highest number of citations) papers.
It can be compared with \tabl{topranked0}, which shows
 {\em top-ranked} papers (papers with highest PaperRank, namely citations weighted proportionally to the rank $R$ of citing papers):
the PaperRank retrieves old famous papers with relatively few citations.
When applied to the time-ordered citation network, the
PageRank reduces to a (weighted) counting of citations-of-citations described in \secc{Rtheo}.
Thereby, when restricted to recent papers, the PaperRank is dominated by the number of individual citations.
Next, \tabl{topreferred0} shows
{\em top-referred} papers, where citations are weighted proportionally to the all-time AuthorRanks $\mathscr{R}_{A}$ of citing authors.
The AuthorRank identifies the recent papers that most attract the attention of notable older authors. Finally the left panel of \fig{Correlations} shows the correlations among paper metrics, showing how our proposed metrics are fairly independent from the traditional ones and among each others.

\medskip

In \secc{AuthorRank} we apply the new metrics to authors.
Traditional metrics shown in \tabl{Npap} are dominated by experimentalists who write more than 100 papers per year in collaborations with more than 2000 authors. Considering instead the number of individual citations, the list in \tabl{Icit} 
becomes dominated by theorists, especially those very active in relatively recent times. 
Restricting to recent papers, the list includes some authors from fields that tend 
produce many publications (tens per author per year). 
Less surprisingly, the list includes authors who produce useful tools for collider experiments, which are presently very active.
Ranking authors trough their PaperRank, 
the all-time list in \tabl{PR} is dominated by theorists
(such as Weinberg, Schwinger, Feynman, Gell Mann) that produced seminal papers after \inspire started,
despite the overall rate of papers and citations was a factor of few smaller than now (\fig{arxivtrends}).
On the other hand, the recent-time list does not change significantly:
the PaperRank is strongly correlated to the number of individual citations, becoming a better metric only on longer time-scales.

Due to this limitation, we developed the AuthorRank. \Tabl{AR} shows the result: 
authors such as Dirac and Einstein now appear in the top of the list, despite having few papers with few citations.
The right columns of \tabl{AR} shows recent authors listed weighting
citations according to the all-time AuthorRank of citing authors.

\Tabl{CCA} lists authors according to their Citation-coin $\CC$:
this metric rewards authors who attract the interest of others by writing above-average papers,
and penalising those that write many below-average papers (or many recent papers,
as papers need decades to be recognized in terms of number of citations).
The right columns of \tabl{CCA} again restrict the list to recent times.

All our lists of authors also show the medals and prizes received by the various authors:
this shows that the non-traditional metrics agree much better 
than traditional metrics with the opinions of the various panels (of course an unknown correlation between bibliometrics and prize awards may still be present).

The right panel of \fig{Correlations} shows the correlations among indicators for authors. 
The metrics we propose are fairly uncorrelated with traditional metrics and among each other.

\medskip

Our metrics respect sum rules (their total is not inflated adding more authors or more references)
and are intensive: this means that groups can be ranked summing over their members.
In \secc{inst} we discussed the institutions that contain the authors that most contributed.
In \secc{towns} we grouped nearby institutes, providing maps of towns most active in fundamental physics in different subfields.
The same is done in \secc{country} for countries \via{and continents}:
in view of the large statistics we also show the time evolution of their percentage impact.
Finally, in \secc{journals} we compute which journals publish the most impactful results in fundamental
physics, again showing the time evolution.

\medskip

The different metrics that we propose give different informations on each author,
providing together a more complete view.
\refchange{{\sc Paperscape}~\cite{paperscape}}{{\sc Paperscape}\footnote{\href{http://paperscape.org}{http://paperscape.org}.}} extracts information from arXiv and provides
a very useful visualisation of the citation graph among papers,
and of the contribution of some authors (those with unique names).
It would be interesting to run the open-source {\sc Paperscape} code on the graph of individual citations among papers and authors extracted from the \inspire database. Our indices could also be implemented in databases that index citations, such as {\sc InSpire}, in order to offer authors (institutes/journals/group) profiles with a larger variety of information, able to give at one glance a much deeper and wider panoramic of each author (institute/journal/group).

Several of our results with complete tables are available at the PhysRank \refchange{webpage~\cite{webpage}}{webpage\footnote{\href{http://rtorre.web.cern.ch/rtorre/PhysRank}{http://rtorre.web.cern.ch/rtorre/PhysRank}.}}.

\medskip

\paragraph{Disclaimer}
Technical details and limitations are described in the appendix.
We repeat the main caveats of our analysis: `fundamental physics' here means `as included in the \inspire database';
we do not correct for mistakes in  \inspire (anyhow more accurate than commercial databases).
Omissions should be addressed to \href{mailto:feedback@inspirehep.net}{feedback@inspirehep.net} or 
trough the on-line forms on \href{https://inspirehep.net}{\sc InSpire}; we will update our results in some future.
We just computed and showed results, avoiding comments.
We hope that authors of any field, journals boards, members of institutes, towns, countries\via{, and continents} will understand that we cannot repeat all caveats in all results.

\subsection*{Note added: 2021 update}
All data have been updated to 2021/1/1.
This is especially interesting, as many activities switched to on-line mode
in view of the  covid-19 pandemic started during early 2020.
Its effect is in progress and will be better evaluated in the future.
So far, \fig{arxivtrends} shows no noticeable impact in the number of publications and
related bibliometric indicators within the main sub-fields.
Similarly, \fig{birthdeathrates} shows no noticeable impact on the number of authors who left the various sub-fields, nor in the number of new authors.
\fig{countries} indicates an accelerated growth of publications from authors with Chinese affiliations 
relative to others.
\fig{continent}a shows that Asia overcome NorthAmerica and that the European contribution, while still larger,
declined significantly. 
This is reflected in \fig{InstEvo}, that shows the time evolution of the percentage impact of
some main institutions.


\small

\subsection*{Acknowledgments}
We thank Roberto Franceschini, Christoffer Petersson, and Paolo Rossi for discussions that stimulated this work.
We thank Roberto Franceschini for participation in the first stage of this project.
We are grateful to the \inspire team, and especially to Jacopo Notarstefano, for support and help with the \inspire database.
\normalsize

\appendix

\section{The \inspire and arXiv databases}\label{app:InSpire}
The open-source \inspire bibliographic database\refchange{~\cite{inspirewebpage}}{\footnote{\href{https://inspirehep.net}{https://inspirehep.net}.}} covers fundamental physics world-wide. 
\sloppy{\inspire presently} contains about $1.4\cdot 10^6$ papers, $4\cdot 10^7$ references, $ 10^5$ authors,
$10^4$ institutions, and $2\cdot 10^{3}$ journals. \inspire started around 1965, but it also contains some notable older papers.
\inspire maps papers, authors, collaborations, institutes (affiliations), and journals to record IDs (integer numbers) thereby addressing the problem of name disambiguation~\citep{DBLP:journals/corr/GolosovskyS16,MartinMontull:2011tya}.

\begin{figure}[t]
\includegraphics[width=1.03\textwidth]{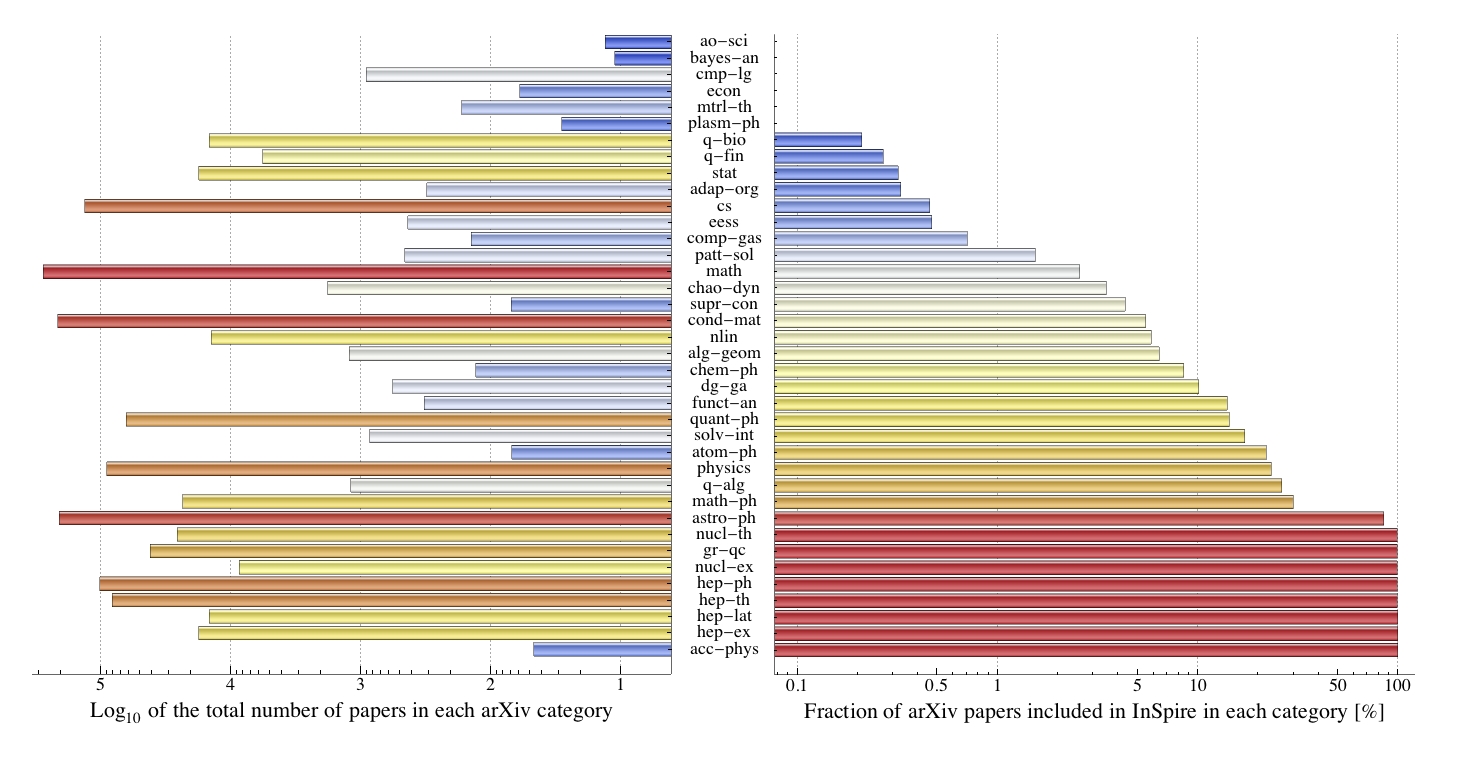}
\caption{\em \label{fig:arXivFraction} Total number of papers in arXiv and fraction included in \inspire by gran categories. The number of papers in each category is computed considering only the main category and ignoring cross-list, so that no paper can belong to more than one gran category. 
\label{fig:arxivcats}}
\end{figure}

Starting from 1995, preprints for most of the papers contained in \inspire are available through arXiv.org\refchange{~\cite{arxivorg},}{,\footnote{\href{https://arxiv.org}{https://arxiv.org}.}} which
also covers fields beyond fundamental physics, so that not all of the arXiv database is included into the \inspire one. The arXiv also provides a classification in terms of categories, some of which contain  sub-classes. The arXiv categories and the number of papers in each of them are shown in the left histogram in \fig{arxivcats}.\footnote{The full list of sub-classes can be found in \refchange{the arXiv API reference manual \cite{arxivapi}}{the arXiv API reference manual (\href{https://arxiv.org/help/api/user-manual}{https://arxiv.org/help/api/user-manual})}.} The right histogram shows the fraction of papers in the various categories included in {\sc InSpire}. 

We also often show results for the main arXiv  categories inside {\sc InSpire}, defined as the arXiv categories with more than $10^{3}$  papers, and with a fraction included in \inspire larger than $50\%$. These are: 
hep-ex (high-energy experiment), hep-ph (high energy theory/phenomenology), 
hep-th (high energy theory),
astro-ph (astrophysics and cosmology),
hep-lat (lattice field theory), 
nucl-ex (nuclear experiment),  nucl-th (nuclear theory), gr-qc (general relativity and quantum cosmology). 
Details of the dataset we consider and technical issues about the \inspire and arXiv databases are discussed in \app{ArXiv1}.

\begin{figure}[t]
$$
\includegraphics[height=0.39\textwidth]{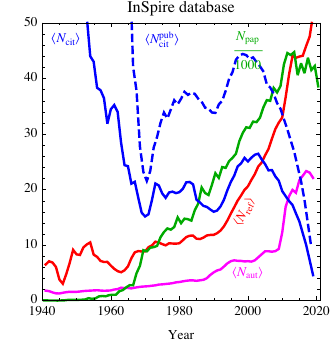}\qquad
\includegraphics[height=0.38\textwidth]{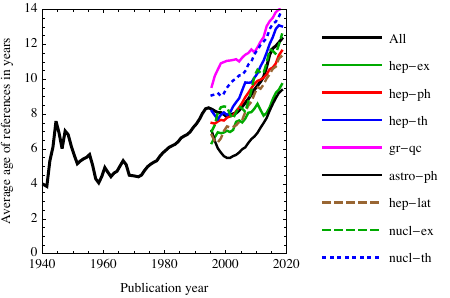}
$$
\caption{\label{fig:trends}\em
{\bf Left}: The vertical axis refers to different quantities:
number of papers per year (green), average number of references (red), of authors (magenta), of citations (blue),
of citations among published papers (blue dashed).
\Fig{arxivtrends} shows the same trends within arXiv categories.
{\bf Right}: average age of references.
Results before $\sim 1960$ are affected by the
smaller size of the literature, and by the incomplete coverage in the \inspire{} database.}
\end{figure}

\begin{figure}[t]
$$\includegraphics[width=\textwidth]{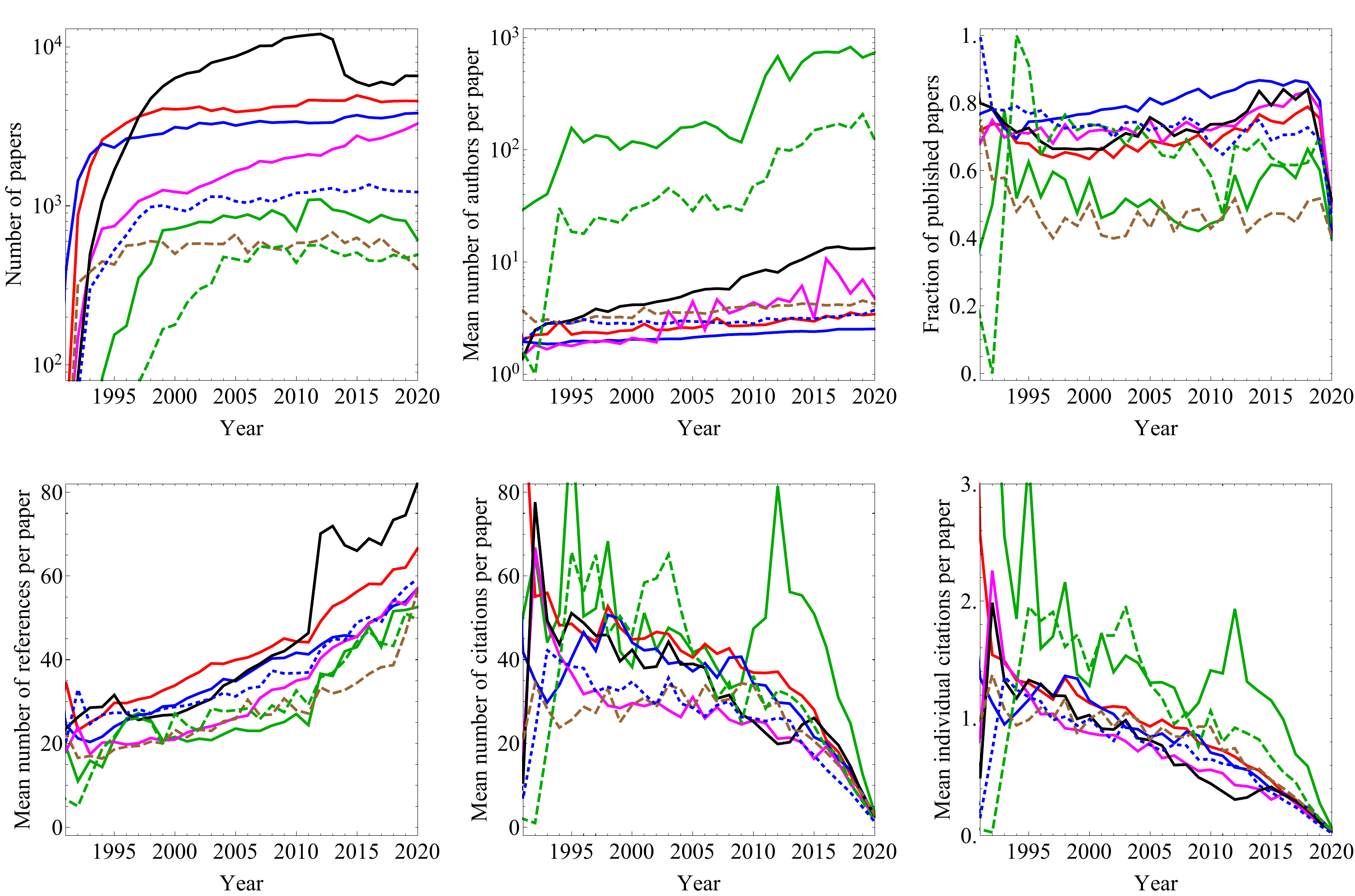}$$
\vspace{-1cm}
\caption{\em \label{fig:arXiv}Trends within the main arXiv representative categories inside \inspire (papers included in \inspire only).
\label{fig:arxivtrends}}
\end{figure}

\begin{figure}[t]
$$\includegraphics[width=\textwidth]{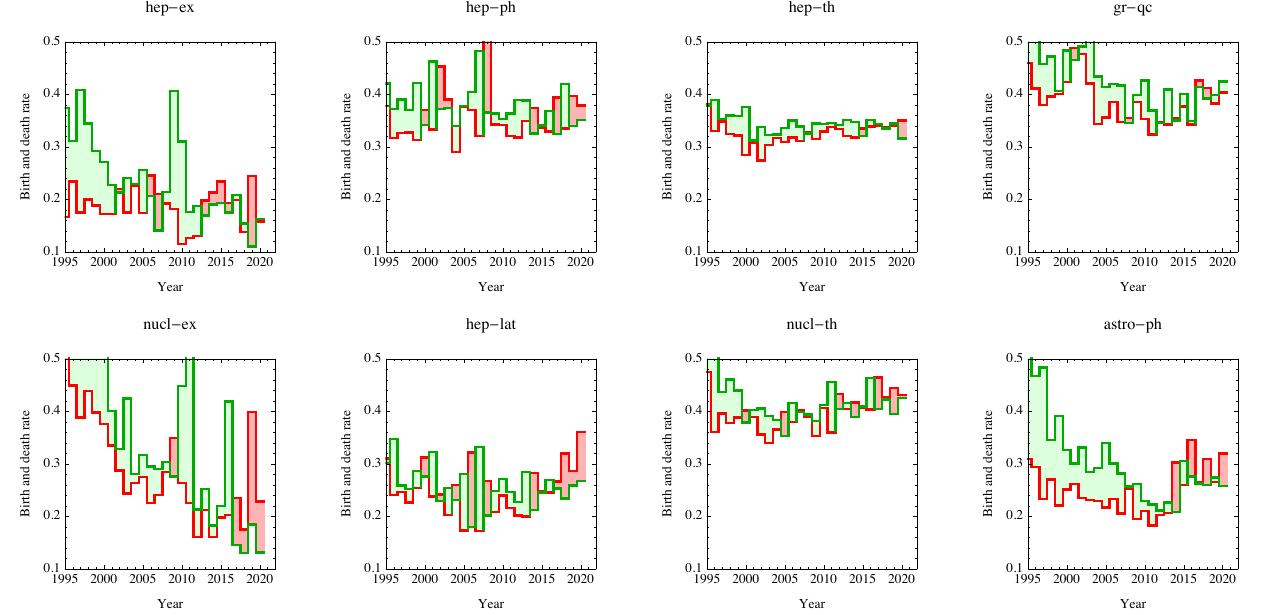}$$
\caption{\em \label{fig:birthdeathrates} Percentage of authors appeared or disappeared each year
within the arXiv categories fully covered by {\sc InSpire}.
}
\end{figure}

\subsection{Main trends in the fundamental physics literature}
The left panel of \fig{trends} shows the time evolution of some main factors:
number of papers per year (which increased by $5\%/{\rm yr}$);
average number of references per paper (increased by $3\%/{\rm yr}$);
number of citations per paper (roughly constant, taking into account
that recent papers, published in the past $\approx 15$ years, necessarily received less citations);
number of authors per paper (increased from few to tens).
We also see that most citations go to published papers. 
In the right panel of \fig{trends} we see that the average age of references is increasing:
before 1980 authors usually mostly cited recent papers.
Contemporary papers cite references published $\Delta t$ earlier
with a distribution $\exp(-\Delta t/\tau)$ with $\tau \approx 11\,{\rm yr}$,
see also \refchange{\reffs{Walker:2006pv,Wang_2013}}{\cite{Walker:2006pv} and \cite{Wang_2013}}.

\smallskip

\Fig{arxivtrends} shows the same trends within the main arXiv representative categories,
showing that papers with an increasingly large number of authors
lie in experimental categories (hep-ex, nucl-ex, astro-ph),
while papers in other fields  keep having, on average, $2-3$ authors.

\smallskip

\Fig{birthdeathrates} shows the authors' `birth' and `death' rates within the main arXiv categories as function
of time: in green the percentage of authors who published in year $y$ but not in year $y-1$;
in red the percentage of authors who published in year $y-1$ but not in year $y$.
The balance is stable, with a significant growth of hep-ex when the Large Hadron Collider (LHC) started, and
of astro-ph until 2010.

\smallskip

\begin{figure}[t]
$$\includegraphics[width=0.5\textwidth]{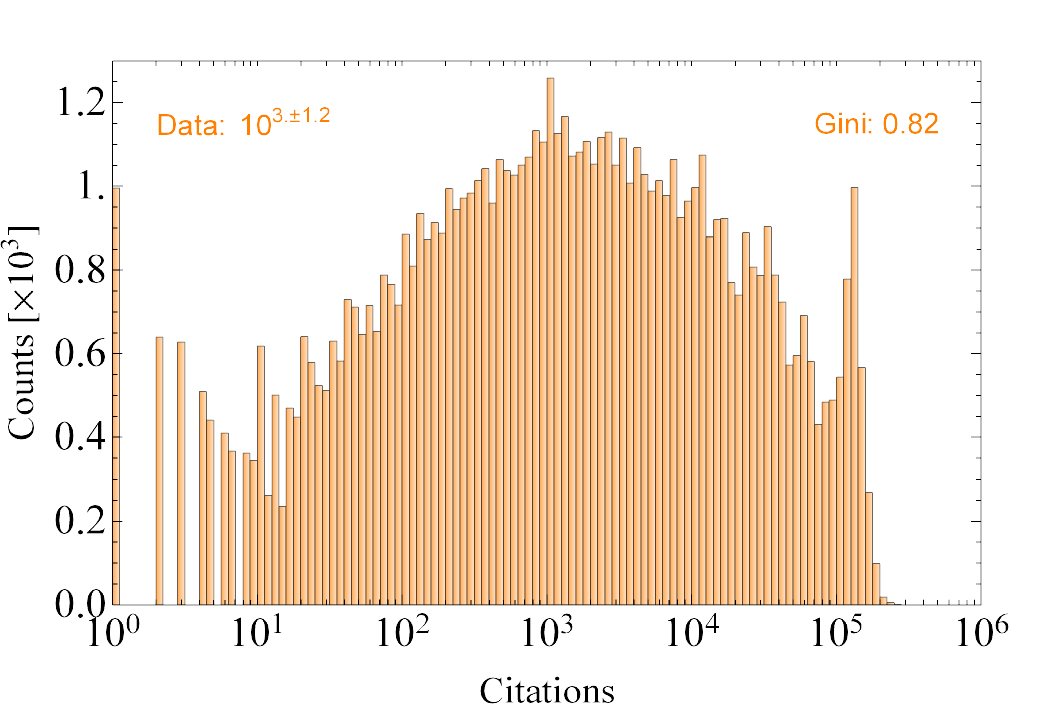}\quad \includegraphics[width=0.5\textwidth]{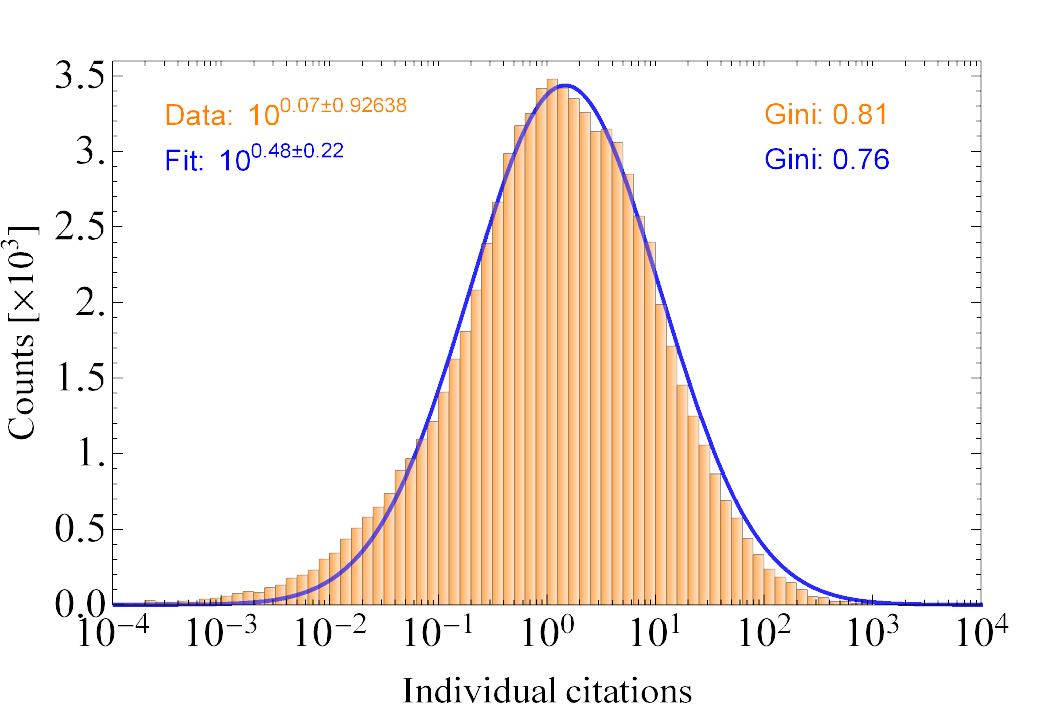}$$
\caption{\em \label{fig:citsdist}
Distribution of citations (left) and individual citations (right) in the whole \inspire database. The blue curve in the right panel is the fit to data with a log-normal distribution with the parameters indicated in the caption. Both the distributions give a Gini index close to 0.8.
}
\end{figure}

\Fig{citsdist} shows the distributions of citations and individual citations for the whole \inspire database. Citations do not follow a well behaved probability distribution due to the presence, in the whole \inspire database, of communities with very different average number of authors. This effect is clearer when going to the individual citations, that apply fractional counting to the citations. Indeed, individual citations are reasonably well described by a log-normal distribution, as already observed \refchange{in \reff{THELWALL2014963}}{by \cite{THELWALL2014963}}, with mean and standard deviation both of order one. This shows how individual citations are well described by a multiplicative stochastic process, allowing to combine and compare more heterogeneous sub-fields. 

In order to give an idea of how impact, as defined from citations and individual citations, is distributed within the community, we computed the Gini index of these distributions.
The Gini index is used in economy as a measure of inequality of wealth. Just to give an idea of its meaning, typical occidental countries presently have Gini coefficients between 0.2 and 0.4. Even though this should not be interpreted analogously to the Gini coefficient in economy, it is interesting to notice that the typical Gini coefficient of citations is between 0.7 (when looking at single arXiv categories) and 0.8 (when considering the whole database). This  means that few papers get most of the citation impact: 
4\% of papers have more than 100 citations, and they receive half of the total citations;
half of the papers have less than 4 citations, and they receive 2\% of the total citations (see also \refchange{\reff{Lehmann:2002tw}}{\cite{Lehmann:2002tw}}).

\subsection{Details about the dataset}\label{app:Database}\label{app:ArXiv1}
Any large database contains a small fraction of incomplete/inconsistent information, which may affect any algorithmic study of the data in a variety of ways \citep{Corso2016AMA}.
The \inspire database is extremely curated.
References are covered with an accuracy at the  $\%$ level, typically better than big private databases such as Scopus\refchange{~\cite{scopus}}{\footnote{\href{https://www.scopus.com}{https://www.scopus.com}.}} or WebOfScience\refchange{~\cite{webofscience}}{\footnote{\href{https://apps.webofknowledge.com}{https://apps.webofknowledge.com}.}}, and comparable to Google Scholar\refchange{~\cite{googlescholar}}{\footnote{\href{https://scholar.google.com}{https://scholar.google.com}.}}.

We obtained the \inspire database in the form of a `dump' file\refchange{ form \reff{inspiredump}.}{.\footnote{\href{http://inspirehep.net/dumps/inspire-dump.html}{http://inspirehep.net/dumps/inspire-dump.html}.}}
\inspire maps papers, authors, collaborations, institutes (affiliations), and journals to record IDs (integer numbers) thereby addressing the problem of name disambiguation~\citep{DBLP:journals/corr/GolosovskyS16,MartinMontull:2011tya}.

The extraction of an accurate date for each paper from the database suffers from some uncertainty. 
There are several available dates: the date when the paper was added to the database, a preprint date, often, but not always, corresponding to the arXiv preprint (when available), a publication date (if the paper has been published on a journal), and an ``earliest'' date, representing the first available date (not always present). Moreover, month information is typically available only for arXiv papers and some published ones. In general we estimate our uncertainty on the extracted dates at the percent level, in the sense that dates are accurately extracted (at least the year) for about 99$\%$ of the papers.
Given this uncertainty our sample of papers consists of 1403881 papers from 1230~\citep{deSacrobosco:1230uu} to 31 December 2020.

\smallskip

Another source of uncertainty comes from the author list of each paper. Some papers carry an empty list of authors. This can have different reasons. For instance, only the name of the collaboration is available
for experimental papers (mainly conference notes) indexed from the CDS database\refchange{~\cite{cerncds}.}{.\footnote{\href{https://cds.cern.ch}{https://cds.cern.ch}.}} All these papers are included in our analysis for what concerns papers, but do not contribute to the metrics for authors. Similar problems extends to institutions and journals.
 
\smallskip
One more source of uncertainty is the extraction of references from papers, needed to produce a citation map. This can be very simple when a bibtex or xml bibliography file is attached to the paper, but can become an extremely complicated task for papers where only a pdf, 
sometimes produced from a scanned paper, is present. \inspire uses state-of-the-art technology for reference extraction\refchange{~\cite{refextractgithub,grobidgithub}}{ (like Refextract\footnote{\href{https://github.com/inspirehep/refextract}{https://github.com/inspirehep/refextract}.} and Grobid\footnote{\href{https://github.com/kermitt2/grobid}{https://github.com/kermitt2/grobid}.})}, which is mainly automatic, with human supervision only in case of errors and inconsistencies. Despite the advanced technology for reference extraction, not all references are correctly extracted. 

There are different kind of problems so that a reference can:
 simply be missed,
 be recognised incompletely,
 be misidentified with another one,
 be assigned to an inexistent paper ID,
 point out of the database (in which case it is counted in the number of references, but not indexed), or
 point to a later paper  (``a-causal'' reference).
All these effects are observed in the database. While some are simple mistakes, or typos in the ID of the paper, the last could be a real effect (below $1\%$), with the reference appearing in a subsequent version of the paper with no available information on the dates of the different versions. Since ``a-causal'' citations can generate anomalies in the computation of the rank (typically only when the a-causality is large, i.e.~in case of mistakes), we deleted them from our dataset. However, since dates are extracted with an accuracy that is often of one year, we still consider causal all the references to papers with the same date, regardless of the actual papers appearing order.
 
In any case, especially because of references pointing outside the database, the number of references indexed for any given record is a better estimate of the actual number of references than the one obtained by summing the indexed ones. Only when computing PaperRanks and AuthorRanks, the number of references is defined equal to the number of indexed references, which is needed to correctly normalize the transition matrix defining the citation graph.
Given that \inspire is complete only after 1970, 
this means that references of older papers are over-attributed to those old notable papers that happen to be in {\sc InSpire}.

In summary, the dataset consists of 1403881 papers, 78941 indexed authors, 8321 institutes, 2644 journals,
37172002 references (of which 26333765 indexed). We compute citations directly from references and do not use citation information from the \inspire database.

Concerning the information on the arXiv database used in \fig{arXivFraction}, we imported all records using the arXiv API\refchange{~\cite{arxivapi}.}{.\footnote{\href{https://arxiv.org/help/api/user-manual}{https://arxiv.org/help/api/user-manual}.}} All other information on arXiv papers and categories has been obtained from the \inspire database. The full list of arXiv categories and subject-classes can be found in the arXiv API reference manual.

All indices discussed in this paper can be computed in one hour on a laptop, apart for the Author Rank, which involves a large, not very sparse matrix: $\sim 8\cdot 10^{4}\times8\cdot 10^{4}$ with about $5\times 10^{8}$ non-vanishing entries.

\footnotesize
\bibliography{artRank}

\end{document}